\documentclass[11pt, draftcls, onecolumn]{./IEEEtran}

\makeatletter
\let\NAT@parse\undefined
\makeatother

\usepackage{epsfig,color}
\usepackage{amsmath,mathrsfs}
\usepackage{amssymb,amsfonts}
\usepackage{amsthm} 
\usepackage[english]{babel}
\usepackage[latin1]{inputenc}
\usepackage{graphicx,color}
\usepackage{amscd}
\usepackage{verbatim}
\usepackage{float}
\usepackage{dsfont}
\usepackage{./citesort}
\usepackage[square, comma]{natbib}

\DeclareMathOperator*{\argmin}{argmin}

\DeclareMathOperator*{\av}{av}
\DeclareMathOperator*{\tr}{tr}
\DeclareMathOperator*{\Det}{det}

\DeclareMathOperator*{\Isom}{Isom}
\DeclareMathOperator*{\Cov}{Cov}
\DeclareMathOperator*{\erfc}{erfc}
\DeclareMathOperator{\arccosh}{arccosh} 
\providecommand{\abs}[1]{\ensuremath{\left\lvert #1 \right\rvert}}
\providecommand{\norm}[1]{\ensuremath{\left\Vert #1 \right\Vert}}

\providecommand{\vv}[1]{\textquotedblleft #1\textquotedblright}
\newcommand{\Q}{\mathbb{Q}}
\newcommand{\Z}{\mathbb{Z}}
\newcommand{\C}{\mathbb{C}}
\newcommand{\R}{\mathbb{R}}

\newcommand{\Hh}{\mathbb{H}}

\newcommand{\ba}{\bar{\alpha}}
\newcommand{\bt}{\bar{\theta}}
\newcommand{\mo}{\mathcal{O}}
\newcommand{\ma}{\mathcal{A}}

\renewcommand{\IEEEQED}{\IEEEQEDopen}

\providecommand{\SL}{SL}

\providecommand{\PSL}{PSL}
\DeclareMathOperator*{\Vol}{Vol}
\DeclareMathOperator*{\n}{N}

\DeclareMathOperator*{\dB}{dB}
\providecommand{\abs}[1]{\ensuremath{\left\lvert #1 \right\rvert}}
\providecommand{\norm}[1]{\ensuremath{\left\Vert #1 \right\Vert}}

\providecommand{\vv}[1]{\textquotedblleft #1\textquotedblright}

\newtheorem{theo}{Theorem}
\newtheorem{lem}[theo]{Lemma}
\newtheorem{prop}[theo]{Proposition}

\newtheorem*{notn}{Notation}
\theoremstyle{definition}
\newtheorem{defn}{Definition}
\newtheorem*{ex}{Example}
\newtheorem{rem}{Remark}

\begin{document}
\title{Algebraic reduction for space-time codes \\ based on quaternion algebras}
\author{L. Luzzi $\qquad$ G. Rekaya-Ben Othman* $\qquad$ J.-C. Belfiore%
\thanks{Jean-Claude Belfiore, Ghaya Rekaya-Ben Othman and Laura Luzzi are with TELECOM ParisTech, 46 Rue\ %
Barrault, 75013 Paris, France. E-mail: $\tt \{belfiore,rekaya,luzzi\}$@$\tt telecom-paristech.fr$. Tel: +33 (0)145817705, +33 (0)145817633, +33 (0)145817636. Fax: +33 (0)145804036 \ %
}}
\bibliographystyle{IEEE}

\maketitle

\begin{abstract}
In this paper we introduce a new right preprocessing method for the decoding of $2 \times 2$ algebraic STBCs, called \emph{algebraic reduction}, which exploits the multiplicative structure of the code. The principle of the new reduction is to absorb part of the channel into the code, by approximating the channel matrix with an element of the maximal order of the algebra.\\
We prove that algebraic reduction attains the receive diversity when followed by a simple ZF detection. Simulation results for the Golden Code show that using MMSE-GDFE left preprocessing, algebraic reduction with simple ZF detection has a loss of only $3 \dB$ with respect to ML decoding.
\medskip \par
\textbf{EDICS category:} MSP-DECD
\end{abstract}

\section{Introduction}
Space-time coding for multiple antenna systems is an efficient device to compensate the effects of fading in wireless channels through diversity techniques, and allows for increased data rates. \\
A new generation of space-time code designs for MIMO channels, based on suitable subsets of division algebras, has been recently developed \cite{SRS}. The algebraic constructions guarantee that these codes are full-rank, full-rate and information-lossless, and have the non-vanishing determinant property.\\
Up to now, the decoding of algebraic space-time codes has been performed using their lattice point representation. In particular, maximum likelihood decoders such as the Sphere Decoder or the Schnorr-Euchner algorithm are currently employed. However, the complexity of these decoders is prohibitive for practical implementation, especially for lattices of high dimension, arising from MIMO systems with a large number of transmit and receive antennas.\\
On the other side, suboptimal decoders like ZF, DFE, MMSE have low complexity but their performance is poor; in particular they don't preserve the diversity order of the system. \\
The use of preprocessing before decoding improves the performance of suboptimal decoders, and reduces considerably the complexity of ML decoders \cite{MEDC}.  Two types of preprocessing are possible:
\begin{itemize}
\item[-] \emph{Left preprocessing} (MMSE-GDFE) to obtain a better conditioned channel matrix;
\item[-] \emph{Right preprocessing} (lattice reduction) in order to have a quasi-orthogonal lattice. The most widely used lattice reduction is the LLL reduction. 
\end{itemize}
We are interested here in the right preprocessing stage; we propose a new reduction method for $2 \times 2$ space-time codes based on quaternion algebras which directly exploits the multiplicative structure of the space-time code in addition to the lattice structure. Up to now, algebraic tools have been used exclusively for coding but never for decoding. \emph{Algebraic reduction} consists in absorbing a part of the channel into the code. This is done by approximating the channel matrix with a unit of a maximal order of the quaternion algebra.\\
The algebraic reduction has already been implemented by Rekaya et al. \cite{RBV} for the fast fading channel, in the case of rotated constellations based on algebraic number fields. In this context, the units in the ring of integers of the field form an abelian multiplicative group whose generators are described by Dirichlet's unit theorem \cite{La}. The reduction algorithm then amounts to decoding in the \emph{logarithmic lattice} of the unit group, which is fixed once and for all.
In this case, one can show that the diversity of the channel is preserved.\\
For quaternion skewfields, which are the object of this paper, the situation is more complicated because the unit group is not commutative. However, it is still possible to find a finite presentation of the group, that is a finite set of generators and relations. \\
Supposing that a presentation is known, we describe an algorithm to find the best approximation of the channel matrix as a product of the generators.\\
As an example, we consider the Golden Code, and find a set of generators for the unit group of its maximal order. Our simulation results for the Golden Code show that using MMSE-GDFE left preprocessing, the performance of algebraic reduction with ZF decoding is within $3 \dB$ of the ML.
\medskip \par
The paper is organized as follows: in Section \ref{system_model} we introduce the system model; in Section \ref{reduction} we explain the general method of algebraic reduction. In Section \ref{section_algorithm} we present the search algorithm to approximate the channel matrix with a unit; in Section \ref{performance} we prove that our method yields diversity order equal to $2$ when followed by a simple ZF decoder. We discuss its performance obtained through simulations in the case of the Golden Code, and compare algebraic reduction and LLL reduction using various decoders (ZF, ZF-DFE), with and without MMSE-GDFE preprocessing.
Finally, in Section \ref{generators_method} we describe a method to obtain the generators of the unit group for a maximal order in a quaternion algebra. The computations are carried out in detail for the case of the algebra of the Golden Code.

\section{System model and notation} \label{system_model}
\subsection{System model}
We consider a quasi-static $2\times2$ MIMO system employing a space-time block code. The received signal is given by
\begin{equation} \label{channel}
Y=HX+W, \qquad X, H, Y, W \in M_2(\C)
\end{equation}
The entries of $H$ are i.i.d. complex Gaussian random variables with zero mean and variance per real dimension equal to $\frac{1}{2}$, and $W$ is the Gaussian noise with i.i.d. entries of zero mean and variance $N_0$.\\
$X$ is the transmitted codeword. In this paper we are interested in STBCs that are subsets of a principal ideal $\mathcal{O}\alpha$ of a maximal order $\mathcal{O}$ in a cyclic division algebra $\mathcal{A}$ of index $2$ over $\Q(i)$ (a quaternion algebra).
We refer to \cite{SRS} for the necessary background about space-time codes from cyclic division algebras, and to \cite{HLRV} for a discussion of codes based on maximal orders. 
\begin{ex}[\textbf{The Golden Code}]
The Golden Code falls into this category (see \cite{BRV} and \cite{HLRV}). It is based on the cyclic algebra $\mathcal{A}=(\Q(i,\theta)/\Q(i),\sigma,i)$, where $\theta=\frac{\sqrt{5}+1}{2}$ and $\sigma: x \mapsto \bar{x}$ is such that $\sigma(\theta)=\bar{\theta}=1-\theta$ and $\sigma$ leaves the elements of $\Q(i)$ fixed.\\
It has been shown in \cite{HLRV} that
\begin{equation} \label{O}
\mathcal{O}=\left\{\left(\begin{array}{cc} x_1 & x_2 \\ i\bar{x}_2 & \bar{x}_1\end{array}\right) ,\; x_1, x_2 \in \Z[i,\theta] \right\}
\end{equation}
is a maximal order of $\mathcal{A}$. $\mathcal{O}$ can be written as $\mo=\Z[i,\theta] \oplus \Z[i,\theta] j$, where 
\begin{equation} \label{j}
j=\left(\begin{array}{cc} 0 & 1 \\ i & 0 \end{array}\right)
\end{equation}
Up to a scaling constant, the Golden Code is a subset of the two-sided ideal $\mathcal{O}\alpha=\alpha\mathcal{O}$, with $\alpha=1+i\bar{\theta}$ \cite{LRBV}. Every codeword of $\mathcal{G}$ has the form  
\begin{equation*}  
X=\frac{1}{\sqrt{5}}\left(\begin{array}{cc} \alpha x_1 & \alpha x_2 \\ \ba i \overline{x}_2  & \ba \overline{x}_1 \end{array}\right)
\end{equation*}
with $x_1=s_1+ s_2 \theta$, $x_2=s_3+s_4 \theta$. The symbols $s_1,s_2,s_3,s_4$ belong to a QAM constellation.
\end{ex}

\subsection{Notation}
In the following paragraphs we will often pass from the $2 \times 2$ matrix notation for the transmitted and received signals to their lattice point representation as complex vectors of length $4$. To avoid confusion, $4 \times 4$ matrices and vectors of length $4$ are written in boldface (using capital letters and small letters respectively), while $2 \times 2$ matrices are not in bold. 

\begin{notn}[\textbf{Vectorization of matrices}]
Let $\phi$ be the function $M_2(\C) \to \C^4$ that vectorizes matrices:
\begin{equation} \label{phi}
\phi: \begin{pmatrix} a & c \\ b & d \end{pmatrix} \mapsto (a,b,c,d)^t
\end{equation}
The left multiplication function $A_l: M_2(\C) \to M_2(\C)$ that maps $B$ to $A B$ induces a linear mapping $\mathbf{A}_l= \phi \circ A_l \circ \phi^{-1} : \C^4 \to \C^4$. That is, 
$$\phi(AB)=\mathbf{A}_l\phi(B) \quad \forall A,B \in M_2(\C)$$
$\mathbf{A}_l$ is the block diagonal matrix 
\begin{equation} \label{left}
\mathbf{A}_l=\begin{pmatrix} A & 0 \\ 0 & A \end{pmatrix}
\end{equation}
\end{notn}

\begin{notn}[\textbf{Lattice point representation}]
Let $\{w_1,w_2,w_3,w_4\}$ be a basis of $\alpha\mo$ as a $\Z[i]$-module. Every codeword $X$ can be written as
$$X=\sum_{i=1}^4 s_i w_i, \quad \mathbf{s}=(s_1,s_2,s_3,s_4)^t \in \Z[i]^4$$
Let $\mathbf{\Phi}$ be the matrix whose columns are 
\begin{equation} \label{Phi}
\phi(w_1),\phi(w_2),\phi(w_3),\phi(w_4)
\end{equation} 
Then the lattice point corresponding to $X$ is
$$\mathbf{x}=\phi(X)=\sum_{i=1}^4 s_i \phi(w_i)=\mathbf{\Phi}\mathbf{s}$$ 
We denote by $\Lambda$ the $\Z[i]$-lattice with generator matrix $\mathbf{\Phi}$. 
\end{notn}

A complex matrix $\mathbf{T}$ is called \emph{unimodular} if the elements of $\mathbf{T}$ belong to $\Z[i]$ and $\det(\mathbf{T}) \in \{1,-1,i,-i\}$. Recall that two generator matrices $\mathbf{\Phi}$ and $\mathbf{\Phi'}$ span the same $\Z[i]$-lattice if $\mathbf{\Phi'}=\mathbf{\Phi}\mathbf{T}$ with $\mathbf{T}$ unimodular. \\
The following remark explains the relation between the units of the maximal order $\mo$ of the code algebra and unimodular transformations of the code lattice. This property is fundamental for algebraic reduction. 

\begin{rem}[\textbf{Units and unimodular transformations}] \label{unit_remark}
Suppose that $U \in \mathcal{O}^*$ is an invertible element: then 
$\{Uw_1,Uw_2,Uw_3,Uw_4\}$ is still a basis of $\alpha\mathcal{O}$ seen as a $\Z[i]$-lattice. The codeword $X$ can also be expressed in the new basis:
$$X=\sum_{i=1}^4 s_i' (U w_i), \quad \mathbf{s'}=(s_1',s_2',s_3',s_4')^t \in \Z[i]^4$$
The vectorized signal is
\begin{equation*}
\mathbf{\Phi} \mathbf{s}=\phi(X)=\sum_{i=1}^4 \phi(Uw_i)=\sum_{i=1}^4 s_i' \mathbf{U}_l \phi(w_i)=\mathbf{U}_l \sum_{i=1}^4 s_i' \phi(w_i)=\mathbf{U}_l \mathbf{\Phi} \mathbf{s}'
\end{equation*}

Now consider the change of coordinates matrix $\mathbf{T}_U=\mathbf{\Phi}^{-1}\mathbf{U}_l \mathbf{\Phi} \in M_4(\C)$ from the basis $\{\phi(w_i)\}_{i=1,\ldots,4}$ to $\{\phi(Uw_i)\}_{i=1,\ldots,4}$. We have 
$\det(\mathbf{T}_U)=\det(\mathbf{U}_l)=\det(U)^2=\pm1,$ 
see equation (\ref{left}). Moreover, we have seen that $\forall \mathbf{s} \in \Z[i]^4$, $\mathbf{s'}=\mathbf{T}_U\mathbf{s} \in \Z[i]^4$. Then $\mathbf{T}_U$ is unimodular, and the lattice generated by $\mathbf{\Phi}\mathbf{T}_U$ is still $\Lambda$. 
\end{rem}

\section{Algebraic reduction} \label{reduction}
In this section we introduce the principle of algebraic reduction. First of all, we consider a normalization of the received signal. In the system model (\ref{channel}), the channel matrix $H$ has nonzero determinant with probability $1$, and so it can be rewritten as
$$H=\sqrt{\det(H)}H_1, \quad H_1 \in \SL_2(\C)$$
Therefore the system is equivalent to
\begin{equation*} 
Y_1=\frac{Y}{\sqrt{\det(H)}}=H_1X+W_1
\end{equation*}
Algebraic reduction consists in approximating the normalized channel matrix $H_1$ with a unit $U$ of norm $1$ of the maximal order $\mathcal{O}$ of the algebra of the considered STBC, that is an element $U$ of $\mathcal{O}$ such that $\det(U)=1$. 

\subsection{Perfect approximation}
In order to simplify the exposition, we first consider the ideal case where we have a perfect approximation: $H_1=U$. Of course this is extremely unlikely in practice; the general case will be described in the next paragraph.\\
The received signal can be written:  
\begin{equation} \label{eq1}
Y_1=UX+W_1
\end{equation}
and $UX$ is still a codeword. In fact, since $U$ is invertible, 
$$\{UX \;|\; X \in \mathcal{O}\alpha\}=\mathcal{O}\alpha$$
Applying $\phi$ to both sides of equation (\ref{eq1}), we find that the equivalent system in vectorized form is 
$$\mathbf{y}_1=\mathbf{U}_l \mathbf{\Phi} \mathbf{s}+\mathbf{w}_1$$
where $\mathbf{\Phi}$ is the matrix defined in (\ref{Phi}), $\mathbf{s} \in \Z[i]^4$, $\mathbf{y}_1=\phi(Y_1)$, $\mathbf{w}_1=\phi(W_1)$.\\
We have seen in Remark \ref{unit_remark} that since $U$ is a unit,
$$\mathbf{U}_l \mathbf{\Phi}=\mathbf{\Phi}\mathbf{T}_U,$$
with $\mathbf{T}_U$ unimodular. So
$$ \mathbf{y}_1=\mathbf{\Phi} \mathbf{T}_U \mathbf{s} + \mathbf{w}_1=\mathbf{\Phi}\mathbf{s_1} + \mathbf{w}_1, \quad \mathbf{s_1} \in \Z[i]^4$$
In order to decode, we can simply consider ZF detection: 
\begin{equation*} 
\mathbf{\hat{s}_1}=\left[\mathbf{\Phi}^{-1}\mathbf{y}_1\right]=\left[\mathbf{s_1}+\frac{1}{\sqrt{\det(H)}}\mathbf{\Phi}^{-1}\mathbf{w}\right]
\end{equation*}
where $[\;]$ denotes the rounding of each vector component to the nearest (Gaussian) integer.\\
If $\mathbf{\Phi}$ is unitary, as in the case of the Golden Code, algebraic reduction followed by ZF detection gives optimal (ML) performance.

\subsection{General case}
In the general case, the approximation is not perfect with probability $1$ and we must take into account the approximation error $E$. We write $H_1=EU$, and the vectorized received signal is 
$$\mathbf{y}_1=\mathbf{E}_l \mathbf{U}_l \mathbf{\Phi} \mathbf{s} + \mathbf{w}_1=\mathbf{E}_l \mathbf{\Phi} \mathbf{T}_U \mathbf{s} + \mathbf{w}_1=\mathbf{E}_l \mathbf{\Phi} \mathbf{s_1} + \mathbf{w}_1$$
The estimated signal after ZF detection is
\begin{equation} \label{decoding}
\mathbf{\hat{s}_1}=\left[\mathbf{\Phi}^{-1}\mathbf{E}_l^{-1} \mathbf{y}_1\right]=\left[\mathbf{s_1}+\frac{1}{\sqrt{\det(H)}}\mathbf{\Phi}^{-1}\mathbf{E}_l^{-1} \mathbf{w}\right]=\left[ \mathbf{s}_1+ \mathbf{n}\right]
\end{equation}
Finally, one can recover an estimate of the initial signal $\mathbf{\hat{s}}=\mathbf{T}_U^{-1}\mathbf{\hat{s}_1}$.\\
Thus, the system is equivalent to a non-fading system where the noise $\mathbf{n}$ is no longer white Gaussian. 

\subsection{Choice of $U$ for the ZF decoder}
We suppose here for simplicity that the generator matrix $\mathbf{\Phi}$ is unitary, but a similar criterion can be established in a more general case. We have seen that ideally the error term $E$ should be unitary in order to have optimality for the ZF decoder, so we should choose the unit $U$ in such a way that $E=H_1U^{-1}$ is quasi-orthogonal. We require that the Frobenius norm $\norm{E}_F^2$ should be minimized\footnote{Remark that since $\det(E)=1$, $\norm{E}_F^2=\norm{E^{-1}}_F^2$.}:
\begin{equation} \label{criterion}
U=\argmin_{\substack{U \in \mathcal{O},\\ \det(U)=1}} \norm{UH_1^{-1}}_F^2
\end{equation}
This criterion corresponds to minimizing the trace of the covariance matrix of the new noise $\mathbf{n}$ in (\ref{decoding}):  
\begin{align*}
&\Cov(\mathbf{n})=\Cov\left(\frac{1}{\sqrt{\det(H)}}\mathbf{\Phi}^{-1}\mathbf{E}_l^{-1} \mathbf{w}\right)=\frac{1}{\abs{\det(H)}} \mathbf{\Phi}^{-1} \mathbf{E}_l^{-1} \Cov(\mathbf{w}) \left(\mathbf{E}_l^{-1}\right)^H \left(\mathbf{\Phi}^{-1}\right)^H=\\
&=\frac{N_0}{\abs{\det(H)}} \mathbf{\Phi}^{-1} \mathbf{E}_l^{-1}\left(\mathbf{E}_l^{-1}\right)^H \left(\mathbf{\Phi}^{-1}\right)^H
\end{align*}
and
\begin{equation} \label{trace}
\tr(\Cov(\mathbf{n})) =\frac{N_0}{\abs{\det(H)}} \norm{\mathbf{\Phi}^{-1}\mathbf{E}_l^{-1}}_F^2 = \frac{N_0}{\abs{\det(H)}} \norm{\mathbf{E}_l^{-1}}_F^2= \frac{2N_0}{\abs{\det(H)}} \norm{E^{-1}}_F^2
\end{equation} 

\section{The approximation algorithm} \label{section_algorithm}

In this section we describe an algorithm to find the nearest unit $U$ to the normalized channel matrix $H_1$ with respect to the criterion (\ref{criterion}). To do this we need to understand the structure of the group of units of the maximal order $\mathcal{O}$.

\begin{notn} We denote elements of $\SL_2(\C)$ with capital letters (for example $H_1,U$) when considering their matrix representation, and with small letters (for example $h_1,u$) when we want to stress that they are group elements. 
\end{notn}

\begin{rem}[\textbf{Units of norm $1$}]
The set 
$$\mo^1=\{u \in \mo^* \;|\; \det(u)=1\}$$
is a subgroup of $\mathcal{O}$. \\
In fact, if $u$ is a unit of the $\Z[i]$-order $\mathcal{O}$, then $\n_{\mathcal{A}/\Q(i)}(u)=\Det(u)$ is a unit in $\Z[i]$, that is, $\Det(u) \in \{1,-1,i,-i\}$. $\mo^1$ is the kernel of the reduced norm mapping $\n=\n_{\mathcal{A}/\Q(i)}:\mathcal{O}^* \to \{1,-1,i,-i\}$ which is a group homomorphism, thus it is a subgroup of $\mo$. 
\end{rem}

\begin{ex}[\textbf{The Golden Code}]
In the case of the Golden Code, $\n$ is surjective since $\n(1)=1$, $\n(\theta)=\theta\bt=-1$, $\n(j)=-j^2=-i$, $\n(j\theta)=i$. So $\{1,-1,i,-i\} \cong \mo^*/\mo^1$, and $\mo^1$ is a normal subgroup of index $4$ of $\mo^*$. In order to obtain a set of generators, it is then sufficient to study the structure of $\mo^1$. Its cosets can be obtained by multiplying for one of the coset leaders $\{1,\theta,j,\theta j\}$. 
\end{ex}

Our problem is then reduced to studying the subgroup $\mo^1$. In particular, we need to find a \emph{presentation} of this group: a set of generators $S$ and a set of relations $R$ among these generators. In fact, one can show that $\mo^1$ is $\emph{finitely presentable}$, that is it admits a presentation with $S$ and $R$ finite \cite{Kl}.

\begin{ex}[\textbf{Generators and relations in the case of the Golden Code}] 
The group $\mo^1$ is generated by $8$ units, that are displayed in Table \ref{generators}. The corresponding relations are shown in Table \ref{relations}.
\end{ex}

The method for finding a presentation is based on the Swan algorithm \cite{Sw}. As it is not well known and is rather complex, we have chosen to expose it in detail in Section \ref{generators_method}.

\begin{table}[btp] 
\begin{center}
\caption{Generators of $\mo^1$.}
\framebox{
$\begin{array}{ll}
u_1=\begin{pmatrix} i\theta & 0 \\ 0 & i\bt \end{pmatrix}=i\theta, &u_1^{-1}=\begin{pmatrix} i\bt & 0 \\ 0 & i\theta \end{pmatrix}=i\bt   \\
u_2=\begin{pmatrix} i & 1+i \\ i-1 & i \end{pmatrix}=i+(1+i)j  &u_2^{-1}=\begin{pmatrix} i & -1-i \\ -i+1 & i \end{pmatrix}=i-(1+i)j   \\
u_3=\begin{pmatrix} \theta & 1+i \\ i-1 & \bt \end{pmatrix}=\theta+(1+i)j  &u_3^{-1}=\begin{pmatrix} \bt & -1-i \\ -i+1 & \theta \end{pmatrix}=\bt-(1+i)j   \\
u_4=\begin{pmatrix} \theta & -1-i \\ -i+1 & \bt \end{pmatrix}=\theta-(1+i)j &u_4^{-1}=\begin{pmatrix} \bt & 1+i \\ i-1 & \theta \end{pmatrix}=\bt+(1+i)j   \\
u_5=\begin{pmatrix} 1+i & 1+i\bt \\ i(1+i\theta) & 1+i \end{pmatrix}=(1+i)+(1+i\bt)j  &u_5^{-1}=\begin{pmatrix} 1+i & -1-i\bt \\ -i(1+i\theta) & 1+i\end{pmatrix}=(1+i)+(1+i\bt)j  \\
u_6=\begin{pmatrix} 1+i & 1+i\theta \\ i(1+i\bt) & 1+i \end{pmatrix}=(1+i)+(1+i\theta)j  &u_6^{-1}=\begin{pmatrix} 1+i & -1-i\theta \\ -i(1+i\bt) & 1+i \end{pmatrix}=(1+i)-(1+i\theta)j   \\
u_7=\begin{pmatrix} 1-i & \bt+i \\ i(\theta+i) & 1-i \end{pmatrix}=(1-i)+(\bt+i)j  &u_7^{-1}=\begin{pmatrix} 1-i & -\bt-i \\-i(\theta+i) & 1-i  \end{pmatrix}=(1-i)-(\bt+i)j   \\
u_8=\begin{pmatrix} 1-i & \theta+i \\ i(\bt+i) & 1-i \end{pmatrix}=(1-i)+(\theta+i)j  &u_8^{-1}=\begin{pmatrix} 1-i & -\theta-i \\ -i(\bt+i) & 1-i \end{pmatrix}=(1-i)-(\theta+i)j  \\
\end{array}$}
\label{generators}
\end{center}
\end{table}

\begin{table}[tbp] 
\caption{Fundamental relations among the generators of $\mo^1$.}
\label{relations}
\begin{center}
\framebox{
$\begin{array}{l}
u_3^{3}=-\mathds{1}\\
u_4^{3}=-\mathds{1}\\
(u_2u_1)^3=\mathds{1}\\
(u_2u_1^{-1})^3=\mathds{1}\\
u_6u_3u_7=-\mathds{1}\\
u_6u_7u_4^{-1}=-\mathds{1}\\
u_8u_3u_5=-\mathds{1}\\
u_4^{-1}u_8u_5=-\mathds{1}\\
u_1u_3^{-1}u_1u_4^{-1}=\mathds{1}\\
u_5^{-1}u_2u_5^{-1}u_1^{-1}u_8u_2u_8u_1=\mathds{1}\\
u_6u_2^{-1}u_6u_1u_7^{-1}u_2^{-1}u_7^{-1}u_1^{-1}=\mathds{1}\\
\end{array}$}
\end{center}
\end{table}

\subsection{Action of the group on the hyperbolic space $\Hh^3$}
The search algorithm is based on the action of the group on a suitable space. We use the fact that $\mo^1$ is a subgroup of the special linear group $\SL_2(\C)$, and consider the action of $\SL_2(\C)$ on the hyperbolic $3$-space $\Hh^3$ (see for example \cite{EGM} or \cite{MR} for a reference). 

We refer to the upper half-space model of $\Hh^3$:
\begin{equation} \label{half_space}
\Hh^3=\{(z,r) \;|\; z \in \C, \; r \in \R, r>0\}
\end{equation}
$\Hh^3$ can also be seen as a subset of the Hamilton quaternions $\mathcal{H}$: a point $P$ can be written as $(z,r)=z+r\mathbf{j}=x+\mathbf{i}y+r\mathbf{j}$, where $\{1,\mathbf{i},\mathbf{j},\mathbf{k}\}$ is the standard basis of $\mathcal{H}$. 
We endow $\Hh^3$ with the hyperbolic distance $\rho$ such that if $P=z+r\mathbf{j}$, $P'=z'+r'\mathbf{j}$,
$$\cosh\rho(P,P')=1+\frac{d(P,P')^2}{2rr'},$$
where $d(P,P')^2=\abs{z-z'}^2+(r-r')^2$ is the squared Euclidean distance.
The corresponding surface and volume forms on $\Hh^3$ are (\cite{MR}, pp. 48--49)
\begin{align}
&ds=\frac{dx^2+dy^2+dr^2}{r^2},\\
&dv=\frac{dx dy dr}{r^3} \label{dv}
\end{align}
The geodesics with respect to this metric are the (Euclidean) half-circles perpendicular to the plane $\{r=0\}$ and with center on this plane, and the half-lines perpendicular to $\{r=0\}$.
Given a matrix
$$g=\begin{pmatrix} a & b \\ c & d \end{pmatrix} \in {\SL}_{2}(\C),$$ 
its action on a point $P=(z,r)$ is defined as follows:
\begin{equation} \label{action}
g(z,r)=(z^*,r^*),\quad \text{ with } \left\{ \begin{array}{l} z^*=\frac{(az+b)(\bar{c}\bar{z}+\bar{d})+a\bar{c}r^2}{\abs{cz+d}^2+\abs{c}^2r^2},\\
                                                                r^*=\frac{r}{\abs{cz+d}^2+\abs{c}^2r^2}\end{array}\right.
\end{equation}
(Here we denote by $\bar{z}$ the complex conjugate of $z$). \\
The action of $g$ and $-g$ is the same, so there is an induced action of $\PSL_2(\C)=\SL_2(\C)/\{\mathds{1},-\mathds{1}\}$. $\PSL_{2}(\C)$ can be identified with the group $\Isom^+(\Hh^3)$ of orientation-preserving isometries of $\Hh^3$ with respect to the metric defined previously (\cite{MR}, p. 48).\\
All the information we will gain about the group $\mo^1$ will thus be modulo the equivalence relation $g \sim -g$; we denote by $P\mo^1$ its quotient with respect to this relation.\\
Consider the action of $\PSL_2(\C)$ on the special point 
\begin{equation} \label{J}
J=(0,1)=\mathbf{j}
\end{equation}
which has the following nice property (\cite{EGM}, Proposition 1.7):
\begin{equation} \label{frobenius_norm}
\forall g \in {\SL}_{2}(\C), \quad\norm{g}_F^2=2\cosh \rho(J,g(J))
\end{equation}

\begin{rem}
If $g \in U(2)$ is unitary, then $g$ leaves every point of $\Hh^3$ fixed (\cite{EGM}, Proposition 1.1). Then by considering for example the mapping $\PSL_2(\C) \to \mathbb{H}^3$ that sends $g$ to $g(J)$, one can identify $\mathbb{H}^3$ with the quotient space $\PSL_2(\C)/U(2)$. 
\end{rem}

\subsection{The algorithm} \label{algorithm}

We assume here the following fundamental properties, which will be proven in Section \ref{generators_method}:
\begin{enumerate}
\item $\{u(J) \;|\; u \in \mo^1\}$ is a discrete set in $\mathbb{H}^3$.
\item Given a unit $u \in \mo^1$, the set
\begin{equation*} 
\mathcal{P}_u=\{P \in \mathbb{H}^3 \; |\; \rho(P, u(J)) \leq \rho(P,u'(J)) \quad \forall u' \neq u\}
\end{equation*}
is a compact hyperbolic polyhedron with finite volume and finitely many faces. 
\item Two distinct polyhedra $\mathcal{P}_u,\mathcal{P}_{u'}$ can intersect at most in one face; all the polyhedra are isometric, and they cover the whole space $\mathbb{H}^3$, forming a \emph{tiling}. Moreover, if $\mathcal{P}=\mathcal{P}_{\mathds{1}}$ is the polyhedron containing $J$, 
$$\mathcal{P}_{u}=u(\mathcal{P})$$
\item The polyhedra adjacent to $\mathcal{P}$ are given by
\begin{equation} \label{adjacent}
u_1(\mathcal{P}),\ldots,u_r(\mathcal{P}),u_1^{-1}(\mathcal{P}),\ldots,u_r^{-1}(\mathcal{P})
\end{equation}
where $\{u_1,\ldots,u_r\}$ is a minimal set of generators for $\mo^1$.
\end{enumerate} 

As anticipated in Section \ref{reduction}, given the normalized channel matrix $h_1 \in \SL_2(\C)$ we want to find
\begin{equation} \label{approximate}
\hat{u}=\argmin_{u \in \mo^1} \norm{uh_1^{-1}}_F^2 
\end{equation}
But we know from equation (\ref{frobenius_norm}) that
\begin{equation*}
\norm{uh_1^{-1}}_F^2=2\cosh(\rho(J,uh_1^{-1}(J)))=2\cosh(\rho(u^{-1}(J),h_1^{-1}(J))),
\end{equation*} 
since $u$ is an isometry. So the condition (\ref{approximate}) is equivalent to
$$\hat{u}=\argmin_{u \in \mo^1} \rho(u^{-1}(J),h_1^{-1}(J))$$
The point $h_1^{-1}(J)$ is contained in the image $\bar{u}(\mathcal{P})=\mathcal{P}_{\bar{u}}$ of the polyhedron $\mathcal{P}$ for some $\bar{u} \in \mathcal{O}^1$. It follows from the definition of $\mathcal{P}_{\bar{u}}$ that $h_1^{-1}(J)$ is closer to $\bar{u}(J)$ than to any other $u(J)$, $u \in \mathcal{O}^1$. Since all the polyhedra are isometric, 
$$\rho(\bar{u}(J),h_1^{-1}(J)) \leq R_{\max} $$
where $R_{\max}$ is the radius of the smallest (hyperbolic) sphere containing $\mathcal{P}$. Therefore we have the following property: 
\begin{equation} \label{hypothesis}
\norm{uh_1^{-1}}_F^2 \leq C_{\mo}
\end{equation}
\medskip \par
We now go back to the problem of finding a unit $\bar{u}$ such that $h_1^{-1}(J) \in \bar{u}(\mathcal{P})$, given a normalized channel matrix $h_1 \in \SL_2(\C)$. 
Let $u_1,\ldots,u_r$ be the generators of $\mo^{1}$ in (\ref{adjacent}) and $u_{r+1}=u_1^{-1},\ldots,u_{2r}=u_r^{-1}$ their inverses. The neighboring polyhedra of $\mathcal{P}$ are all of the form $u_i(\mathcal{P})$, $i=1,\ldots,2r.$\\
The idea is to begin the search from $\mathcal{P}$ and the neighboring polyhedra, corresponding to the generators of the group and their inverses, and choose the $U_i$ such that $u_i(J)$ is the closest to $h_1^{-1}(J)$. Since $u_i$ is an isometry of $\mathbb{H}^3$, at the next step we can apply $u_i^{-1}$ and start again the search of the $u_{i'}$ that gives the closest point to ${u_i}^{-1}h_1^{-1}(J)$. With this strategy we only need to update a single point and perform $2r$ comparisons at each step of the search.

\begin{figure}[btp] 
\begin{center}
\includegraphics[width=0.8\textwidth,clip=true]{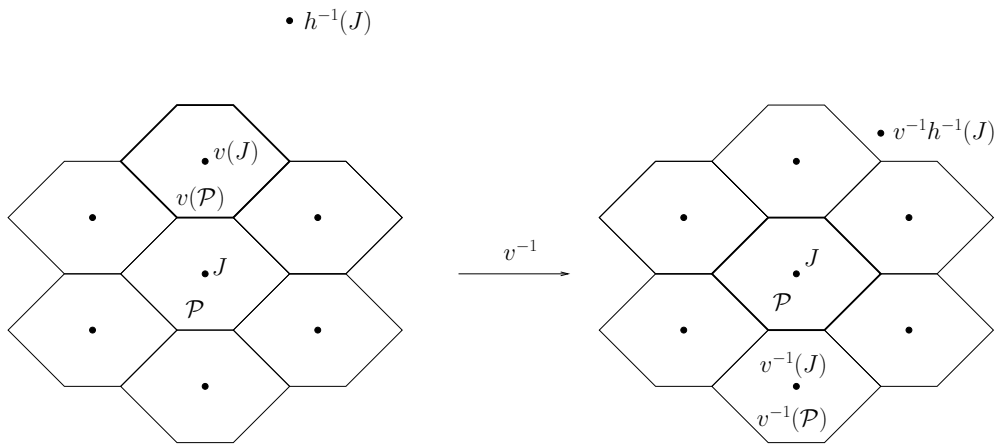}
\caption{A step of the algorithm. The polyhedra are represented as two-dimensional polygons for simplicity.}
\label{polygons}
\end{center}
\end{figure}

The algorithm is illustrated in Figure \ref{polygons}. \\
Suppose that the matrix form of the $u_i$ has been stored in memory at the beginning of the program, together with the images $u_1(J),\ldots,u_{2r}(J)$ of $J$, for example using the coordinates in the upper half-space model (\ref{half_space}). Let 
$$u_i(J)=(x_i,y_i,r_i), \quad i=1,\ldots,2r$$
\textbf{INPUT:} \emph{$h_1 \in \SL_2(\C)$.} \\
\emph{Initialization: let $h=h_1$,$\bar{u}=\mathds{1}$,$i_0=0$.}\\
\textbf{REPEAT} 
\begin{enumerate}
\item \emph{Compute $h^{-1}(J)=(x,y,r)$.} 
\item \emph{Compute the distances} 
{\allowdisplaybreaks
\begin{align*}
&d_i=2\cosh\rho(h^{-1}(J),u_i(J))=1+\frac{(x-x_i)^2+(y-y_i)^2+(r-r_i)^2}{2rr_i}, \quad i=1,\ldots,2r, \\
&d_0=2\cosh\rho(h^{-1}(J),J)
\end{align*}
}
\item \emph{Let $i_0=\argmin_{i \in \{0,1,\ldots,2r\}} d_i$. (If several indices $i$ attain the minimum, choose the smallest.)} 
\item \emph{Update $\bar{u}\leftarrow \bar{u}u_{i_0}$, $h \leftarrow hu_{i_0}$.}
\end{enumerate}
\textbf{UNTIL} $i_0=0$.\\
\textbf{OUTPUT:} \emph{$\hat{u}=\bar{u}^{-1}$ is the chosen unit.}

\begin{rem}[\textbf{Advantage of the algebraic reduction in the case of slow fading channels}]
If the channel varies slowly from one time block to the next, it is reasonable to expect that the polyhedron $\bar{u}(\mathcal{P})$ containing $h_1^{-1}(J)$ at the time $t$ will be the same, or will be adjacent, to the polyhedron chosen at the time $t-1$. Thus, this method requires only a slight adjustment of the previous search at each step. On the contrary, the LLL reduction method requires a full lattice reduction at each time block.
\end{rem}

\section{Performance of the algebraic reduction} \label{performance}

\subsection{Diversity} \label{diversity}

It has recently been proved \cite{TMK} that MIMO decoding based on LLL reduction followed by zero-forcing achieves the receive diversity. The following Proposition shows that algebraic reduction is equivalent to LLL reduction in terms of diversity for the case of $2$ transmit and $2$ receive antennas: 

\begin{prop} \label{prop_diversity}
The diversity order of the algebraic reduction method with ZF detection is $2$.
\end{prop}

\begin{IEEEproof}
We suppose that the symbols $s_i$, $i=1,\ldots,4$ belong to an $M$-QAM constellation, with $M=2^{2m}$. Let $\mathcal{E}_{\av}$ be the average energy per symbol, and $\gamma=\frac{\mathcal{E}_{\av}}{N_0}$ the SNR.\\ 
For a fixed realization of the channel matrix $H$, equation (\ref{decoding}) is equivalent to an additive channel without fading where the noise $\mathbf{n}$ is no longer white. \\
We can compute the error probability using ZF detection conditioned to a certain value of $H$, and then average over the distribution of $H$:
\begin{equation} \label{average_Pe}
P_e(\gamma)=\int P_e(\gamma\;|\;H)dH
\end{equation}
With symbol by symbol ZF detection, $P_e(\gamma)$ is bounded by the error probability for each symbol:
$$P_e(\gamma) \leq \sum_{i=1}^4 P((\mathbf{\hat{s}_1})_i \neq (\mathbf{s_1})_i),$$
Using the classical expression of $P_e$ in a Gaussian channel, for square QAM constellations (\cite{Pr}, \S5.2.9), we obtain
$$P((\mathbf{\hat{s}_1})_i \neq (\mathbf{s_1})_i) \leq 4\erfc\left(\sqrt{\frac{3 \mathcal{E}_{\av}}{\sigma^2(M-1)}}\right) \leq 4e^{-\frac{3\mathcal{E}_{\av}}{2(M-1)\sigma_i^2}}$$
where $\sigma_i^2$ is the variance for complex dimension of the noise component $n_i$ .\\
We have seen in (\ref{trace}) that the trace of the covariance matrix of the new noise $\mathbf{n}$ is bounded by $\frac{N_0}{\abs{\det(H)}} \norm{\mathbf{\Phi^{-1}}}_F^2 \norm{\mathbf{E}_l^{-1}}_F^2$, recalling that the Frobenius norm is submultiplicative. Thus 
\begin{align*}
&\sigma_i^2 \leq \Cov(\mathbf{n}) \leq \frac{CN_0}{\abs{\det(H)}}
\end{align*} 
because $\norm{E^{-1}}_F^2=\norm{UH_1^{-1}}_F^2 \leq C_{\mo}$, see equation (\ref{hypothesis}).\\
Indeed if $\mathbf{\Phi}$ is unitary, as in the case of the Golden Code, $\norm{\mathbf{\Phi}^{-1}\mathbf{E}_l^{-1}}_F^2=\norm{\mathbf{E}_l^{-1}}_F^2 \leq C_{\mo}$. \\ 
Finally,
$$P_e(\gamma\;|\;H) \leq 16 e^{-\left(\frac{3}{2(M-1)C}\right)\abs{\det(H)}\frac{\mathcal{E}_{\av}}{N_0}}=16 e^{-c\abs{\det(H)}\gamma}$$
In order to compute the error probability in equation (\ref{average_Pe}), we need the distribution of $\abs{\det(H)}$. It is known \cite{Go,Ed} that if $H$ is gaussian with i.i.d. $\mathcal{N}(0,1)$ entries (variance per real dimension $\frac{1}{2}$), the random variable $4\abs{\det(H)}^2$, corresponding to the determinant of the \emph{Wishart matrix} $2HH^H$, is distributed as the product of two independent chi square random variables with $2$ and $4$ degrees of freedom respectively.\\
Consider two random variables $X \sim \chi^2(2)$,$Y \sim \chi^2(4)$: their joint probability distribution function is
$$p_{X,Y}(x,y) =\frac{1}{8} y e^{-\frac{x}{2}-\frac{y}{2}} \qquad x,y>0$$ 
Then the cumulative distribution function of $Z=2\abs{\det(H)}=\sqrt{XY}$ is
$$F_Z(z)=P\{\sqrt{XY} \leq z\} =\iint_{\sqrt{xy} \leq z} p_{X,Y} (x,y) dxdy$$
From the invertible change of variables $u=y$, $v=\sqrt{xy}$ with Jacobian
$$J=\left| \begin{array}{cc} -\frac{v^2}{u^2} & \frac{2v}{u} \\ 1 & 0 \end{array} \right|=-\frac{2v}{u}$$
we obtain
\begin{align*}
&F_Z(z)=\int_0^z \int_0^{\infty} p_{X,Y} \left(\frac{v^2}{u},u\right) \abs{J} du dv=\int_0^z \frac{v}{4} \left( \int_0^{\infty} e^{-\frac{v^2}{2u}-\frac{u}{2}} du\right) dv,\\
&p_Z(z)=\frac{\partial F_Z(z)}{\partial z} =\frac{z}{4} \int_0^{\infty} e^{-\frac{z^2}{2u}-\frac{u}{2}} du =\frac{z^2}{2} K_1(z),
\end{align*}
where $K_1$ is the modified Bessel function of the second kind. Finally,
\begin{multline*}
P_e(\gamma) \leq \mathbb{E}\left[16 e^{-c' \gamma Z}\right]=16 \int_0^{\infty} \frac{z^2}{2} K_1(z) e^{-c' \gamma z} dz=\\
=16\left(\frac{1}{(c' \gamma)^2} +\frac{2}{\pi (c'\gamma)^4}\sum_{k=0}^{\infty} \left(\frac{1}{(c'\gamma)^{2k}} \frac{\Gamma(k+\frac{5}{2})}{\Gamma(k+1)}\left(\Psi\left(k+\frac{5}{2}\right)-\Psi(k+1)-2\ln(c' \gamma)\right)\right)\right), 
\end{multline*}

where $\Psi$ is the Digamma function. The series in the last expression being uniformly bounded for large $\gamma$, the leading term is of the order of $\frac{1}{\gamma^2}$.
\end{IEEEproof}

\subsection{Some remarks about complexity}
The length of the algorithm described in Section \ref{algorithm} is related to the initial distance 
\begin{equation*}
2\cosh\rho(h_1^{-1}(J),J)=\norm{h_1^{-1}}_F^2=\norm{h_1}_F^2=\norm{\frac{H}{\sqrt{\det{H}}}}_F^2=\frac{\norm{H}_F^2}{\abs{\det(H)}} 
\end{equation*}
In order to have more information about the distribution of this distance, one has to find the distribution of the random variable $\frac{\norm{H}_F^2}{\abs{\det(H)}}$. From \cite{Ed}, we learn that $H$ is unitarily similar to 
$$\widetilde{H}=\frac{1}{2}\begin{pmatrix} X & 0 \\ Y & Z \end{pmatrix},$$
where $X^2 \sim \chi^2(4)$, $Y^2,Z^2 \sim \chi^2(2)$ and $X,Y,Z$ are independent. Therefore
\begin{equation*}
\frac{\norm{H}_F^2}{\abs{\det(H)}}=\frac{\norm{\widetilde{H}}_F^2}{\abs{\det(\widetilde{H})}} = \frac{X^2+ Y^2 + Z^2}{XZ} 
\end{equation*}
We want to find the distribution of the random variable  $T=\frac{X^2+ Y^2 + Z^2}{XZ}$ knowing the distributions of $X,Y,Z$:
$$p_X(x)=\frac{x^3}{2} e^{-\frac{x^2}{2}},\quad p_Y(y)= y e^{-\frac{y^2}{2}},\quad p_Z(z)=z e^{-\frac{z^2}{2}}$$ 
Their joint probability distribution is
$$p_{X,Y,Z}(x,y,z)=\frac{1}{2} x^3 y z e^{-\frac{1}{2}(x^2+y^2+z^2)},$$
and the distribution of $T$ is given by
\begin{align*}
&p_T(t)=\frac{\partial}{\partial t} F_T(t)=\frac{\partial}{\partial t} \iint_{\frac{x^2+y^2+z^2}{xz} \leq t} p_{X,Y,Z} (x,y,z) dx dy dz=\\
&=\frac{\partial}{\partial t}\int_0^\infty \int_{\frac{x(t-\sqrt{t^2-4})}{2}}^{\frac{x(t+\sqrt{t^2-4})}{2}} \int_0^{\sqrt{txz-x^2-z^2}} \frac{x^3 y z}{2} e^{-\frac{x^2+y^2+z^2}{2}} dy dz dx
\end{align*}

With the change of variables $\frac{t}{2}=\cosh(u)$, $w=y^2$ this integral becomes
\begin{align*}
&\frac{\partial}{\partial u} \left(\int_0^\infty \int_{x(\cosh(u)-\sinh(u))}^{x(\cosh(u)+\sinh(u))} \int_0^{2\cosh(u)xz-x^2-z^2}  \frac{x^3z}{4} e^{-\frac{x^2+z^2+w}{2}}dw dz dx\right) \frac{\partial u}{\partial t}= \\
&=\frac{\partial}{\partial u} \tanh^3(u) \frac{\partial u}{\partial t}=\frac{12\sqrt{t^2-4}}{t^4}
\end{align*}

The following example shows that the distance $\rho(h_1^{-1}(J),J)$ is mostly concentrated near the origin:
\begin{ex}
In Section \ref{generators_method}, we will see that the minimum of the distances between $J$ and the vertices of $\mathcal{P}$ for the Golden Code is $R_{\min}=\arccosh(1.9069\cdots)=1,2614\cdots$. 
From the distribution of $T$, we find that the probability that $\rho(J,h_1^{-1}(J))> R_{\min}$ is approximately $0.038$: in most cases $h_1^{-1}(J)$ is already 
contained in $\mathcal{P}$ or in one of the neighboring polyhedra, and the algorithm stops after one step! Moreover, the probability that $\rho(J,h_1^{-1}(J))>5 R_{\min}$ is of the order of $10^{-10}$, which is negligible since for practical values of the SNR the probabilities of error for ML detection are typically of the order of $10^{-6}$ at best.
\end{ex}


\subsection{Simulation results} \label{simulations}

\begin{figure}[tbp] 
\begin{center}
\includegraphics[width=0.5\textwidth]{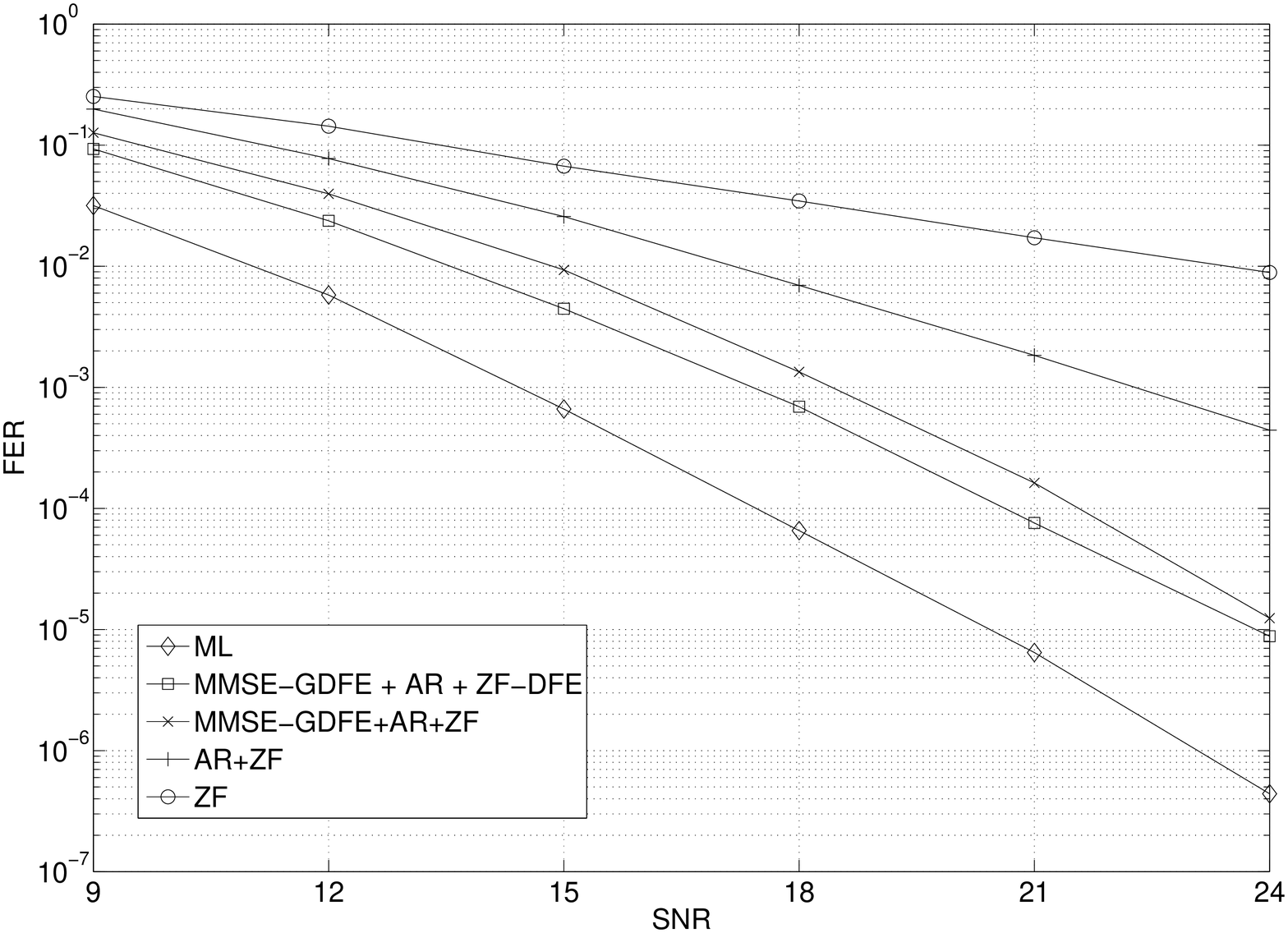}
\caption{Performance of algebraic reduction followed by ZF or ZF-DFE decoders using $4$-QAM constellations.} 
\label{4QAM}
\end{center}
\end{figure}

\begin{figure}[tbp] 
\begin{center}
\includegraphics[width=0.5\textwidth]{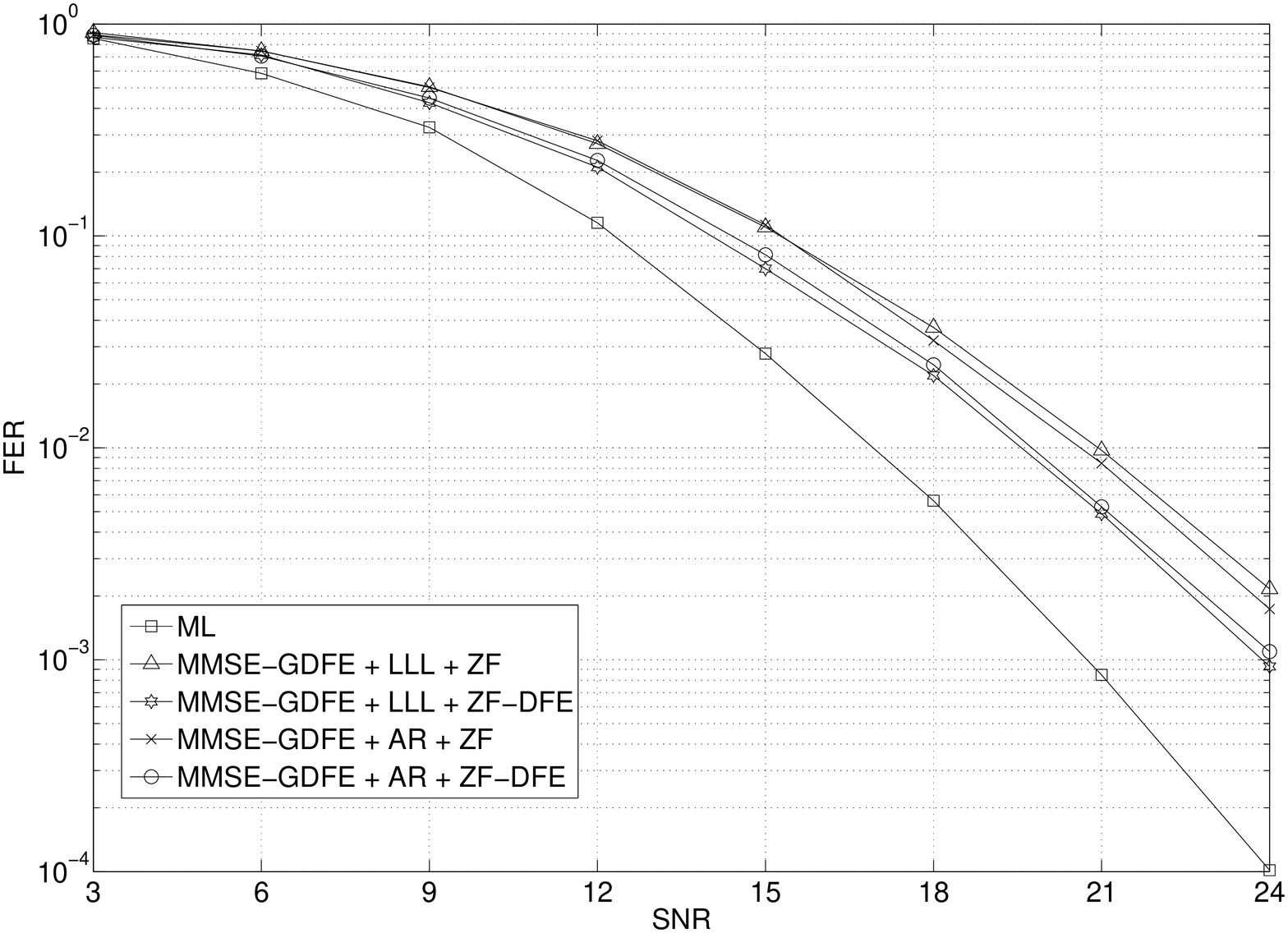}
\caption{Comparison of algebraic reduction and LLL reduction using MMSE-GDFE preprocessing combined with ZF or ZF-DFE decoding with $16$-QAM constellations.} 
\label{16QAM}
\end{center}
\end{figure}

Figure \ref{4QAM} shows the performance of algebraic reduction followed by ZF and ZF-DFE decoding compared with ML decoding using $4$-QAM constellations. One can verify that the slope of the probability of error in the case of algebraic reduction with ZF detection (without preprocessing) is very close to $-2$, confirming the result of Proposition \ref{prop_diversity} concerning the diversity order. \\
One can add MMSE-GDFE left preprocessing to solve the shaping problem for finite constellations \cite{MEDC} in order to improve this performance. With MMSE-GDFE preprocessing, algebraic reduction is within $4.2\dB$ and $3.2\dB$ from the ML using ZF and ZF-DFE decoding, at the FER of $10^{-4}$.\\
In the $16$-QAM case, the loss is of $3.4\dB$ and $2.6\dB$ respectively for ZF and ZF-DFE decoding at the FER of $10^{-3}$ (Figure \ref{16QAM}). In the same figure we compare algebraic reduction to LLL reduction using MMSE-GDFE preprocessing. The two performances are very close; with ZF-DFE decoding, algebraic reduction has a slight loss ($0.3\dB$). On the contrary, with ZF decoding, algebraic reduction is slightly better ($0.4\dB$ gain), showing that the criterion (\ref{criterion}) is indeed appropriate for this decoder. 

\medskip \par
Numerical simulations also evidence that the average complexity of algebraic reduction is low. In Section \ref{algorithm} we have seen that each step of the unit search algorithm requires only a few operations. Table \ref{compl} shows the actual distribution of the number of steps in the unit search algorithm. The data refers to a computer simulation for the Golden Code using a ZF decoder, for 16-QAM constellations, for the transmission of $10^{6}$ codewords. (Clearly this distribution does not depend on the SNR.) The average length of the algorithm is less than $2$.

\begin{table}[btp]
\begin{center}
\begin{tiny}
\caption{Number of steps of the search algorithm.}
\begin{tabular}{|c| |c|c|c|c|c|c|c|c|c|c|c|}
\hline
average number of steps &  \multicolumn{11}{|c|}{distribution of the number of steps} \\
\hline
&  $1$ & $2$ & $3$ & $4$ & $5$ & $6$ & $7$ & $8$ & $9$ & $10$ & $>11$  \\
\hline
$1.923062$ & $38.2\%$ & $39.4\%$ & $16.0\%$ & $4.8\%$ & $1.2\%$ & $0.2\%$ & $3.7 \cdot 10^{-2}\%$ & $7.6 \cdot 10^{-3}\%$ & $2.6 \cdot 10^{-3}\%$ & $10^{-4}\%$ & $0$ \\ 
\hline
\end{tabular}
\label{compl}
\end{tiny}
\end{center}
\end{table}

\section{Finding the generators} \label{generators_method}

In this Section, we describe a method to find a presentation for the group $\mo^1$ of units of norm $1$, and the corresponding polyhedron $\mathcal{P}$. The computations are carried out in detail for the Golden Code. 

\subsection{Kleinian groups and Dirichlet polyhedra}
We introduce some terminology that will be useful later:

\renewcommand{\theenumi}{\alph{enumi}}

\begin{defn}[\textbf{Kleinian groups}]
Let $\Gamma$ be a subgroup of the projective special linear group $\PSL_2(\C)$ acting on the hyperbolic space $\mathbb{H}^3$. 
\begin{itemize}
\item[-] If $\Gamma$ is \emph{discrete}, that is if the subspace topology on $\Gamma$ is the discrete topology, $\Gamma$ is called a \emph{Kleinian group}. Remark that then $\Gamma$ is countable. 
\item[-] The \emph{orbit} of a point $x_0 \in \mathbb{H}^3$ is the set $\{g(x_0) \;|\; g \in \Gamma\}$.
\item[-] A \emph{fundamental set} for the action of $\Gamma$ is a subset of $\Hh^3$ containing exactly one point for every orbit. 
\item[-] A \emph{fundamental domain} for $\Gamma$ is a closed subset $D$ of $\Hh^3$ such that 
\begin{enumerate}
\item $\bigcup_{g \in \Gamma} g(D)=\Hh^3$,
\item If $g \in \Gamma \setminus \{\mathds{1}\}$, the interior of $D$ is disjoint from the interior of $g(D)$.
\item The boundary of $D$ has measure $0$.
\end{enumerate}
\item[-] $\Gamma$ is called \emph{cocompact} if it admits a compact fundamental domain; we say that $\Gamma$ has \emph{finite covolume} if it admits a fundamental domain with finite volume. 
\item[-] If $\Gamma$ has finite covolume, and $D_1$ and $D_2$ are fundamental domains for $\Gamma$, then $\Vol(D_1)=\Vol(D_2)< \infty$ (\cite{MR}, Lemma 1.2.9).
\end{itemize}
\end{defn}

In the case of a Kleinian group $\Gamma$, one can obtain a fundamental domain that is a hyperbolic polyhedron \cite{MR,CJLdR}. This polyhedron can be obtained as an intersection of hyperbolic half-spaces. \\
For any pair of distinct points $Q, Q' \in \Hh^3$, the set of points equidistant to $Q$ and $Q'$ with respect to $\rho$ is a hyperbolic plane, called the \emph{bisector} between $Q$ and $Q'$, which divides $\Hh^3$ into two open convex half-spaces, one containing $Q$ and the other containing $Q'$. Given $g \in \Gamma$, let 
\begin{equation} \label{bisector} 
D_g(Q)=\{P \in \Hh^3\;|\; \rho(Q,P) \leq \rho(g(Q),P)\}
\end{equation}
the closed half-space of the points that are closer to $Q$ than to $g(Q)$.
If $Q$ is not fixed by any nontrivial element of $\Gamma$, the \emph{Dirichlet fundamental polyhedron of $\Gamma$ with center $Q$} is defined as the intersection of all the bisectors corresponding to nontrivial elements: 
\begin{equation} \label{polyhedron}
\mathcal{P}_{\Gamma}=\bigcap_{\substack{g \in \Gamma,\\ g\neq \mathds{1}}} D_g(Q)
\end{equation}
The definition (\ref{polyhedron}) cannot be used directly to compute the polyhedron, since we ought to intersect an infinite number of bisectors. Let  $\overline{B}(Q,R)$ denote the closed ball with center $Q$ and radius $R$, and let 
\begin{equation} \label{ball}
D_R(Q)=\bigcap \{D_g(Q) \;|\; g \neq \mathds{1}, \; g(Q) \in \overline{B}(Q,R)\}
\end{equation}
If $\mathcal{P}_{\Gamma}$ is compact, it has finite diameter, so there exists $\overline{R}>0$ such that $\mathcal{P}_{\Gamma}=D_{\overline{R}}(Q)$.

\subsection{Poincaré's theorem}
From the Dirichlet polyhedron of a Kleinian group one can obtain a complete description of the latter, including generators and relations. In fact, a famous theorem due to Poincaré establishes a correspondence between a set of generators of the group and the isometries which map a face of the polyhedron into another face, called \emph{side-pairings}. The sequences of side-pairings which send an edge into itself correspond to a complete set of relations among the generators. \\
A complete exposition of Poincaré's theorem in the general case can be found in \cite{EP}. We only need a rather weak version of the theorem that we state as follows: 
\begin{theo}[\textbf{Poincaré's polyhedron theorem}] \label{poincare}
Let $\mathcal{P}$ be a hyperbolic polyhedron in $\Hh^3$ with finitely many faces. Let $\mathcal{F}$ denote the set of faces of $\mathcal{P}$, and suppose that:
\begin{enumerate}
\item[a)] \emph{[\textbf{Metric condition}]}\; For every pair of disjoint faces of $\mathcal{P}$, the corresponding geodesic planes have no common point at infinity.
\item[b)] \emph{[\textbf{Side-pairings}]}\; There exist two maps $\mathcal{R}: \mathcal{F} \to \mathcal{F}$, $\mathcal{U}: \mathcal{F} \to \Isom(\Hh^3)$ such that:
\begin{itemize}
\item[-] $\forall F \in \mathcal{F}$, $\mathcal{R}^2(F)=F$
\item[-] If $\mathcal{R}(F)=F'$, $\mathcal{U}(F)=u_F$ maps $F'$ onto $F$, sending distinct vertices into distinct vertices, and distinct faces into distinct faces, and maps the interior of $\mathcal{P}$ outside of $\mathcal{P}$. Moreover $u_{F'}=(u_F)^{-1}$
\end{itemize}
$\mathcal{R}$ is called a \emph{side-pairing} for $\mathcal{P}$.
\item[c)] \emph{[\textbf{Cycles}]}\; For each edge $E_1$ of $\mathcal{P}$, there is a \emph{cycle} starting with $E_1$, that is a sequence of the form $[E_1,\ldots,E_{n+1}]$, where $E_i$, $i=1,\ldots,n+1$ are edges of $\mathcal{P}$, and  $\forall i \in \{1,\ldots,n\}$ there exists a generator $u^{(i)} \in \mathcal{U}(\mathcal{F})$ such that $u^{(i)}(E_i)=E_{i+1}$, and $E_{n+1}=E_1$. Moreover, we suppose that $u=u^{(n)}\circ \cdots \circ u^{(1)}$ is a rotation through an angle $\frac{2\pi}{m}$, $m \in \Z^+$, and that its restriction to $E_1$ is the identity. 
\end{enumerate} 
Consider the group $\Gamma$ generated by $\mathcal{U}(\mathcal{F})$. Then $\mathcal{P}$ is a fundamental domain for the action of $\Gamma$ on $\Hh^3$.
\end{theo}  
The proof of this theorem is a special case of the proof of Theorem 4.14 in \cite{EP}.

\subsection{The structure of $\mo^1$} \label{structure}
We now have all the necessary background to find a fundamental domain, and thus a set of generators, for $P\mo^1=\mo^1/\{\mathds{1},-\mathds{1}\}$. The following theorem shows that $P\mo^1$ is a Kleinian group, and describes its Dirichlet polyhedron (see \cite{EGM} or \cite{Vig}):
\begin{theo} \label{t1}
Let $\ma$ be a quaternion algebra over a number field $K$ such that
\begin{enumerate}
\item[a)] $K$ has exactly one pair of complex embeddings
\item[b)] $\ma$ is ramified at all the real places, that is $\mathcal{A} \otimes_{\Q} K_{\nu}$ is a division ring for every real place $\nu$ of $K$.
\end{enumerate}
Let $\mathcal{O}$ be an order of $\ma$. Then:
\begin{itemize}
\item[-] $P\mo^1$ is a Kleinian group. 
\item[-] $P\mo^1$ has finite covolume and its Dirichlet polyhedron has finitely many faces.
\item[-] $P\mo^1$ is cocompact if and only if $\ma$ is a division ring. 
\end{itemize}
\end{theo}
Remark that conditions (a) and (b) of the theorem are verified since $K=\Q(i)$ is an imaginary quadratic number field and thus has a pair of complex embeddings and no real embeddings. 
\medskip \par
Thus $P\mo^1$ admits a compact fundamental polyhedron $\mathcal{P}_{\mo^1}$ of the form (\ref{polyhedron}), with finitely many faces and finite volume. This volume is known \emph{a priori} and only depends on the choice of the algebra $\ma$ (see \cite{MR}, p.336):
\begin{theo}[\textbf{Tamagawa Volume Formula}] \label{theorem_volume}
Let $\ma$ be a quaternion algebra over $K$ such that $\ma \otimes_{\Q} \R \cong M_2(\C)$. Let $\mo$ be a maximal order of $\ma$. Then the hyperbolic volume
$$\Vol(\mathcal{P}_{\mo^1})=\frac{1}{4\pi^2}\zeta_K(2)\abs{D_K}^{\frac{3}{2}}\prod_{p |\delta_\mathcal{O}} (N_p-1)$$
In the previous formula, $\zeta_K$ denotes the \emph{Dedekind zeta function}\footnote{The Dedekind zeta function is defined as $\zeta_K(s)=\sum\limits_{I} ([O_K:I])^{-s}$, where $I$ varies among the proper ideals of $O_K$.} relative to the field $K$, $D_K$ is the discriminant of $K$, $\delta_{\mathcal{O}}$ is the discriminant of $\mathcal{O}$, $p$ varies among the primes of $O_K$, and $N_p=[O_K:pO_K]$, where $O_K$ is the ring of integers of $K$.
\end{theo}

\begin{ex}[\textbf{The Golden Code}]
In the case of the Golden Code algebra, $D_{\Q(i)}=-4$, and $\delta(\mathcal{O})=5\Z[i]$. The only primes that divide the discriminant of the maximal order are $(2+i)$ and $(2-i)$, both with algebraic norm $5$. In conclusion,
\begin{equation} \label{volume}
\Vol(\mathcal{P}_{\mo^1})=\frac{8\zeta_{\Q(i)}(2)16}{4\pi^2}=\frac{32 \zeta_{\Q(i)}(2)}{\pi^2}=4,885149838\cdots 
\end{equation}
since $\zeta_{\Q(i)}(2)=1.50670301\cdots$.
\end{ex}

\begin{rem} We have seen in section \ref{algorithm} that a smaller polyhedron $\mathcal{P}$ results in a better average distance between $\hat{u}(J)$ and $h^{-1}(J)$ and a better approximation. So the algebraic codes such that $\Vol(\mathcal{P})$ is small are better suited for the method of algebraic reduction.
\end{rem}

\subsection{Computing the Dirichlet polyhedron for $\mo^1$} \label{strategy}
We suppose here that we already have an estimate of the volume of $\mathcal{P}_{\mo^1}$, given by Theorem \ref{theorem_volume}. For this reason our strategy to find the Dirichlet polyhedron differs slightly from that described in \cite{CJLdR}. The idea is to compute the sequence $D_R(Q)$ defined by (\ref{ball}) for an increasing sequence of values of $R$, until we find $\overline{R}$ such that $D_{\overline{R}}(Q)$ is compact. If the hypotheses of Poincaré's Theorem are verified for a set of side-pairings belonging to $\mo^1$, $D_{\overline{R}}(Q)$ is a Dirichlet polyhedron for some subgroup of $\mo^1$. To check whether this subgroup coincides with $\mo^1$ it is sufficient to estimate of the volume of $D_{\overline{R}}(Q)$.
\medskip \par
Following \cite{CJLdR}, we take as our base point $Q$ the point $J$ defined in (\ref{J}). One needs to check that $J$ is not fixed by any nontrivial element of $P\mo^1$.

\begin{rem}
As pointed out in \cite{CJLdR}, if on the contrary $J$ is fixed by some nontrivial element, one needs first to compute a fundamental domain for the stabilizer $\Gamma_J$ of $J$ (the subgroup of elements that fix $J$), and then intersect it with $\bigcap_{\substack{g \in \Gamma \setminus \Gamma_J}} D_g(J)$.
\end{rem}
Because of the property (\ref{frobenius_norm}), in order to find $D_R(J)$ we only need to intersect the bisectors corresponding to elements of $\mo^1$ with square Frobenius norm less or equal to $2\cosh(R)$. 
Since $\mo$ can be identified with a discrete lattice in $\C^4$ using the map $\phi$ defined in (\ref{phi}), and the Frobenius norm corresponds to the Euclidean norm in $\C^4$, clearly there is only a finite number of these elements. 
\medskip \par
Since $\cosh$ is increasing on the positive half-line, in order to find the half-space $D_g(J)$ one can solve the inequality $\cosh(\rho(Q,J)) \leq \cosh(\rho(Q,g(J)))$. For a general $g=\begin{pmatrix} a & b \\ c& d \end{pmatrix}$, 
$$g(J)=\left(\frac{b\bar{d}+a\bar{c}}{\abs{d}^2+\abs{c}^2},\frac{1}{\abs{d}^2+\abs{c}^2}\right),$$
and the corresponding half-space has equation
\begin{equation*}
(C-1)x^2+(C-1)y^2+(C-1)r^2-2Ax-2By+\frac{A^2+B^2+1}{C}-1 \geq 0,
\end{equation*}
where $A=\Re(b\bar{d}+a\bar{c})$,$B=\Im(b\bar{d}+a\bar{c})$, $C=\abs{d}^2+\abs{c}^2$. Its boundary is a sphere of center $\left(\frac{A}{C-1},\frac{B}{C-1},0\right)$ and square radius $\frac{1}{C}\left(\frac{A^2+B^2}{(C-1)^2}+1\right)$.
Remark that if we change the sign of the pair $a,d$ or $b,c$, the radius doesn't change, while the center is reflected with respect to the origin.\\
If a ball or complementary of a ball (according to the sign of $C-1$) in the list (\ref{ball}) is already contained in the intersection, we can discard the corresponding element of the group. Since all the spheres have center on the plane $\{r=0\}$, in order to determine whether a sphere is contained in another we only need to consider their intersections with this plane.

\subsection{Computing the generators for the Golden Code}

We now apply the method described in the previous Section to the maximal order $\mathcal{O}$ of the algebra of the Golden Code. 
One can easily verify that in this case $J$ is not fixed by any nontrivial element of $\mo^1$.\\
Considering the elements $g \in \mo^1$ such that $\norm{g}_F^2 \leq 9$ by computer search, we find that $\mathcal{P}=D_{\overline{R}}(J)$ is compact with $\overline{R}=\arccosh\left(\frac{9}{2}\right)$, since it doesn't intersect the plane \vv{at infinity} $\{r=0\}$. Table \ref{generators} lists the elements $\{u_i,u_i^{-1}\}, i=1,\ldots,8$ of the group that are necessary to obtain $\mathcal{P}$.
The equations of the corresponding spheres, the bisectors $$S(u_i)=D_{u_i}(J), \quad S(u_i^{-1})=D_{u_i^{-1}}(J), \qquad i=1,\ldots,8$$ (see definition (\ref{bisector})) can be found in Table \ref{equations}.

\begin{figure}[tbp]
 \begin{minipage}[b]{.46\linewidth}
\centering\includegraphics[width=\textwidth,height=\textwidth,angle=270]{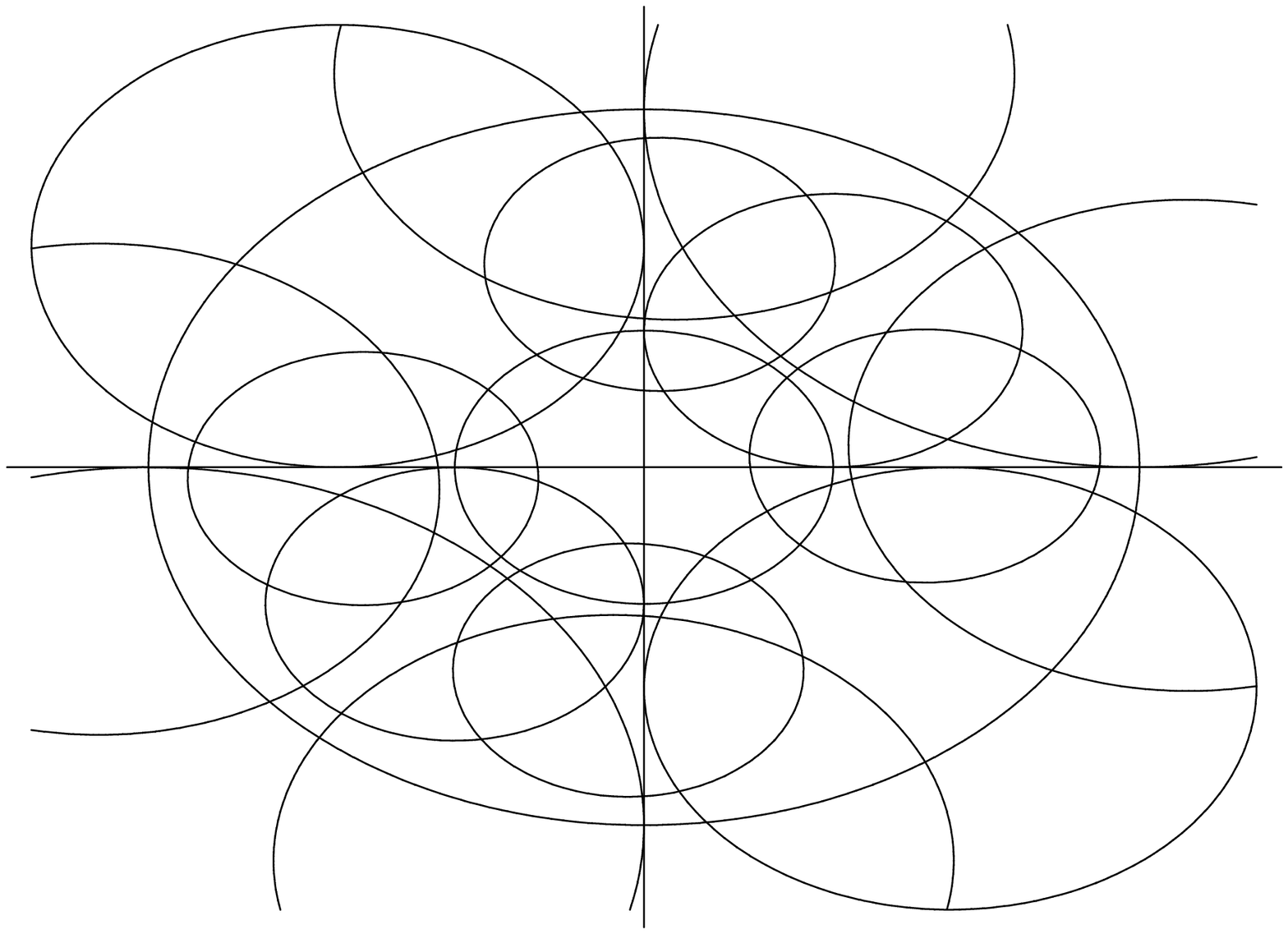}
  \caption{The projection of the bisectors on the plane $\{r=0\}$. \label{circles}}
 \end{minipage} \hfill
 \begin{minipage}[b]{.46\linewidth}
\centering\includegraphics[angle=270,width=\textwidth]{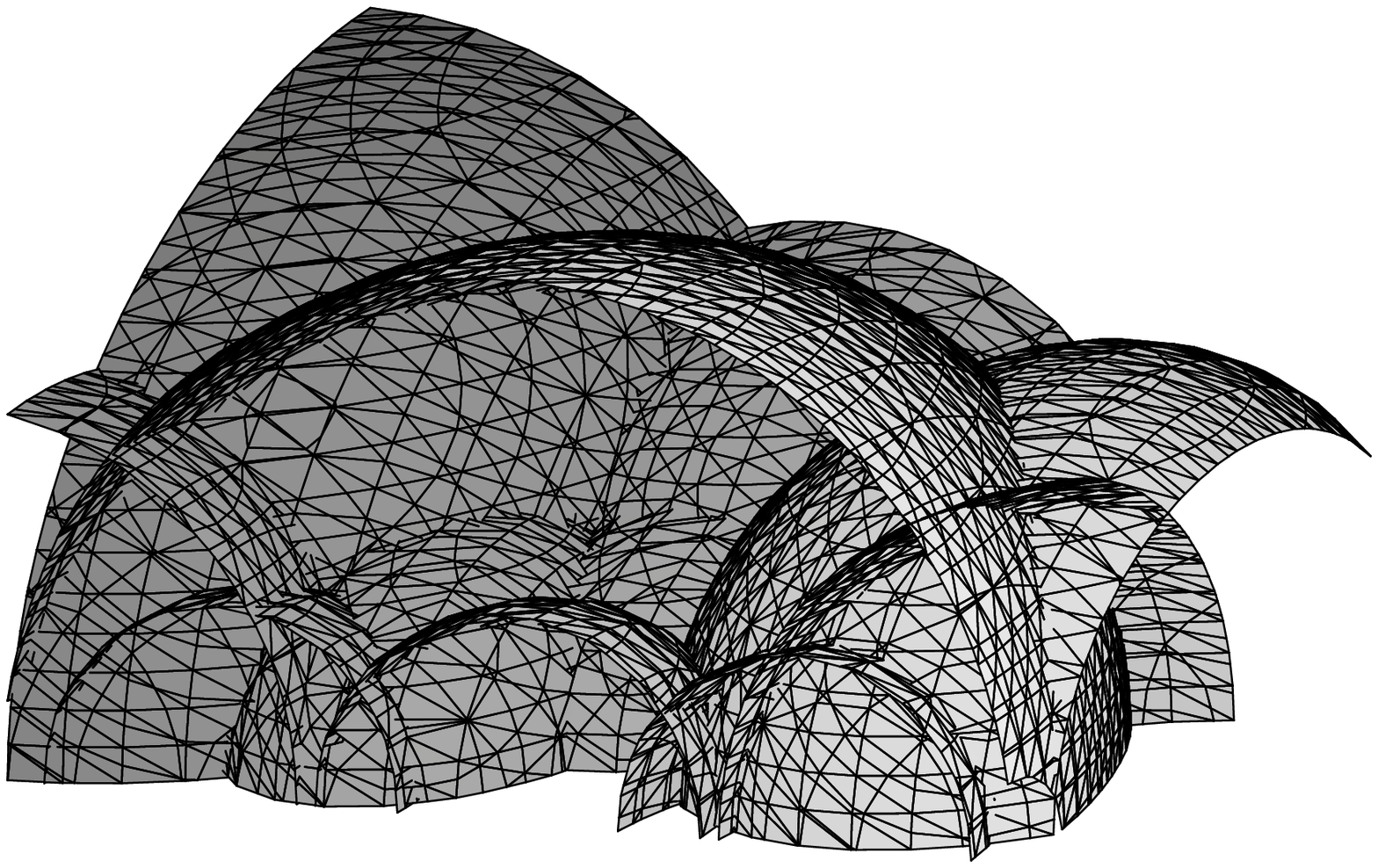}
\caption{The intersection of the spheres (the picture shows only half of the space for better understanding). \label{section}}
 \end{minipage}
\end{figure}



\begin{table}[tbp] \begin{center}
\caption{Bisectors}
\begin{tabular}{|l|c|c|c|}
\hline
unit &  center & radius & int/ext \\
\hline
$u_1$ & $(0,0,0)$ & $\theta$ & I\\
$u_1^{-1}$  & $(0,0,0)$ & $\bt$ & E\\  
$u_2$ &  $(1,1,0)$  & $1$ & E\\
$u_2^{-1}$ &  $(-1,-1,0)$  & $1$ & E \\
$u_3$ & $(-\theta,-\theta,0)$ & $\theta$ & E  \\
$u_3^{-1}$ & $(\bt,\bt,0)$ & $-\bt$ & E \\
$u_4$ & $(\theta,\theta,0)$ & $\theta$ & E \\
$u_4^{-1}$ & $(-\bt,-\bt,0)$ & $-\bt$  & E\\
$u_5$ & $\left(\frac{-9\sqrt{5}+19}{22},\frac{-9-5\sqrt{5}}{22},0\right)$ & $\frac{\sqrt{7}}{22}(7-\sqrt{5})$ & E  \\
$u_5^{-1}$ & $\left(\frac{9\sqrt{5}-19}{22},\frac{9+5\sqrt{5}}{22},0\right)$ & $\frac{\sqrt{7}}{22}(7-\sqrt{5})$ & E \\
$u_6$ & $\left(\frac{9\sqrt{5}+19}{22},\frac{-9+5\sqrt{5}}{22},0\right)$ & $\frac{\sqrt{7}}{22}(7+\sqrt{5})$ & E \\
$u_6^{-1}$ & $\left(\frac{-9\sqrt{5}-19}{22},\frac{9-5\sqrt{5}}{22},0\right)$ & $\frac{\sqrt{7}}{22}(7+\sqrt{5})$ & E\\
$u_7$ & $\left(\frac{-9-5\sqrt{5}}{22},\frac{-9\sqrt{5}+19}{22},0\right)$ & $\frac{\sqrt{7}}{22}(7-\sqrt{5})$ & E\\
$u_7^{-1}$ & $\left(\frac{9+5\sqrt{5}}{22},\frac{9\sqrt{5}-19}{22},0\right)$ & $\frac{\sqrt{7}}{22}(7-\sqrt{5})$ & E\\
$u_8$ & $\left(\frac{-9+5\sqrt{5}}{22},\frac{9\sqrt{5}+19}{22},0\right)$ & $\frac{\sqrt{7}}{22}(7+\sqrt{5})$ & E\\
$u_8^{-1}$ & $\left(\frac{9-5\sqrt{5}}{22},\frac{-9\sqrt{5}-19}{22},0\right)$ & $\frac{\sqrt{7}}{22}(7+\sqrt{5})$ & E \\
\hline
\end{tabular}
\label{equations}
\end{center}
\end{table}

\begin{figure}[tbp]
 \begin{minipage}[b]{.46\linewidth}
\centering\includegraphics[width=.7\textwidth,height=\textwidth,angle=270]{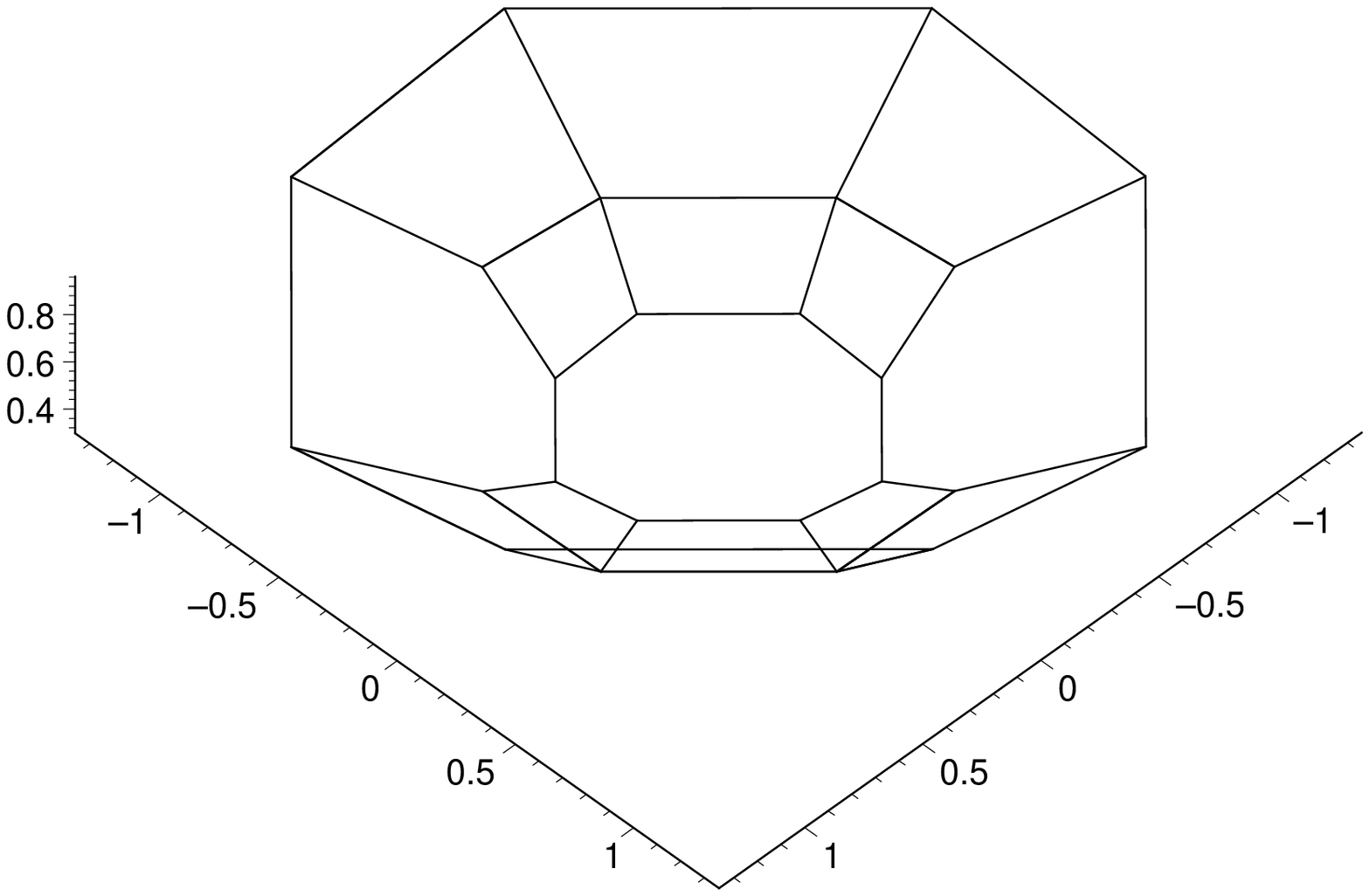}
  \caption{A schematic representation of the polyhedron $\mathcal{P}$ (in the picture, the edges have been replaced by straight lines).}
\label{P}
 \end{minipage} \hfill
 \begin{minipage}[b]{.46\linewidth}
\label{flat}
\centering\includegraphics[width=0.8\textwidth,clip=true]{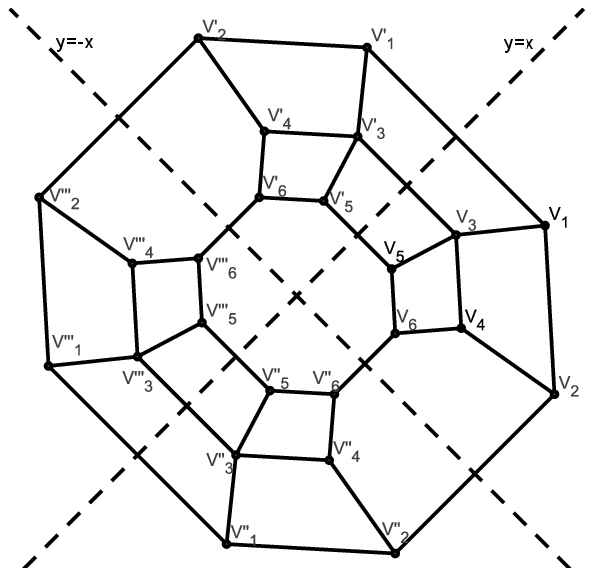}
\caption{The projection of the polyhedron $\mathcal{P}$ on the plane $\{r=0\}$.}
 \end{minipage}
\end{figure}



The vertices of $\mathcal{P}$ are the intersections of all the triples of spheres $S(u_i)$: 
\begin{equation*}
V_i,\;V_i'=\pi'(V_i),\; V_i''=\pi''(V_i),\;V_i'''=\pi'''(V_i),\quad i=1\ldots 6, 
\end{equation*}
Here $\pi'$,$\pi''$ and $\pi'''$ denote the reflections with respect to the plane $\{y=x\}$, the plane $\{y=-x\}$ and the line $\{x=0,y=0\}$ respectively.
{ \allowdisplaybreaks
\begin{align*}
& V_1=\left(\frac{5\sqrt{5}+9}{16},\frac{3\sqrt{5}-1}{16},\frac{1}{8}\sqrt{33+11\sqrt{5}}\right)=S(u_1) \cap S(u_4) \cap S(u_6),\\
& V_2=\left(\frac{1+\theta}{2},-\frac{1}{2},\frac{\theta}{2}\right)=S(u_1)\cap S(u_2) \cap S(u_6),\\
& V_3=\left(\frac{\theta}{2},-\frac{\bt}{2},\frac{1}{2}\right)=S(u_4)\cap S(u_6) \cap S(u_7^{-1}),\\
& V_4=\left(\frac{3\sqrt{5}}{20}+\frac{1}{2},\frac{3\sqrt{5}}{20}-\frac{1}{2},\frac{1}{2}\sqrt{\frac{11}{10}}\right)=S(u_2) \cap S(u_6) \cap S(u_7^{-1}) \\
& V_5=\left(\frac{1+3\sqrt{5}}{16},\frac{5\sqrt{5}-9}{16},\frac{1}{8}\sqrt{33-11\sqrt{5}}\right)=S(u_1^{-1}) \cap S(u_4^{-1}) \cap S(u_7^{-1}),\\
& V_6=\left(\frac{1}{2},-\frac{\bt^2}{2},\frac{\bt}{2}\right)=S(u_1^{-1})\cap S(u_2) \cap S(u_7^{-1})
\end{align*}
}


The faces of $\mathcal{P}$ correspond to portions $F(u)$ of the spheres $S(u)$, with $u$ one of the units in Table \ref{generators}. The projection of the faces of $\mathcal{P}$ on the plane $\{r=0\}$ is shown in Figure \ref{flat}.

As explained in Section \ref{strategy}, in order to prove that $\mathcal{P}$ is a Dirichlet polyhedron for $\mo^1$, we will first show that it is a fundamental domain for some subgroup $\Gamma$ of $\mo^1$ using Poincaré's Theorem. Comparing the volume of $\mathcal{P}$ with the value (\ref{volume}), we will find that $\Gamma=\mo^1$.\\ 
The metric condition in Poincaré's Theorem can be verified given the equations of the spheres (see also Figure \ref{circles}). \\
Define a side-pairing as follows: $\mathcal{U}(F(u))=u$, $\mathcal{R}(F(u))=F(u^{-1})$ for every $u$ in Table \ref{generators}. The action of the generators on the faces and vertices is summarized in Table \ref{action2}, and it is not hard to see that it satisfies all the conditions in the theorem. In fact every face $F(u)=\mathcal{P} \cap u(\mathcal{P}) \subset S(u)$. 
Remark that an isometry between polygons with the same number of vertices, sending distinct vertices in distinct vertices, must be onto. 

In order to check that the cycle condition of Theorem \ref{poincare} holds, we need to compute the minimal relations or \vv{cycles} between the generators, by finding the sequences of edges of $\mathcal{P}$ of the form $[E_1,\ldots,E_{n+1}]$, such that $u^{(i)}(E_i)=E_{i+1}$, and $E_{n+1}=E_1$. As $F((u^{(i)})^{-1})$ must contain $E_{i}$,  there are only two possible choices for $u^{(i)}$, corresponding to the two faces containing the edge $E_i$. \\
Given such a sequence, $u^{(1)}\cdots u^{(n)}$ is an element of finite order in $\mo^1$, that is  $(u^{(1)}\cdots u^{(n)})^k=\mathds{1}$ for some $k$. (Remark that every cyclic permutation of the sequence $[E_1,\ldots,E_{n+1}]$ gives rise to a new cycle.) Actually it is necessary to \vv{lift} the relation from $\PSL_2(\C)$ to $\SL_2(\C)$.  We also require our sequences to be \emph{irreducible}, that is $u^{(i+1)} \neq (u^{(i)})^{-1}$ for all $i$. 
\medskip \par
In this way we obtain a decomposition of the set of edges of $\mathcal{P}$ into cycles. The action of the generators on the faces is summarized in Table \ref{action2}; the cycles are described in Table \ref{cycles}.

\begin{table}[btp] 
\caption{Cycles for the edges of $\mathcal{P}$.}
\label{cycles}
\begin{center}
\begin{tabular}{|l|}
\hline
$\overline{V_3''V_3'''} \xrightarrow{u_3} \overline{V_3''V_3'''}$\\
$\overline{V_3V_3'} \xrightarrow{u_4} \overline{V_3V_3'}$ \\
$ \overline{V_6V_6''} \xrightarrow{u_1} \overline{V_2'''V_2'} \xrightarrow{u_2} \overline{V_6V_6''} $ \\
$ \overline{V_2 V_2''} \xrightarrow{u_1^{-1}} \overline{V_6'''V_6'} \xrightarrow{u_2} \overline{V_2V_2''} $ \\
$ \overline{V_3V_4} \xrightarrow{u_6^{-1}} \overline{V_3'''V_1} \xrightarrow{u_3^{-1}} \overline{V_3'''V_5'''} \xrightarrow{u_7^{-1}} \overline{V_3V_4} $ \\
$ \overline{V_1V_3} \xrightarrow{u_6^{-1}} \overline{V_4'''V_3'''} \xrightarrow{u_7^{-1}} \overline{V_5V_3} \xrightarrow{u_4} \overline{V_1V_3} $ \\
$ \overline{V_3'V_4'} \xrightarrow{u_5} \overline{V_3''V_5''} \xrightarrow{u_3} \overline{V_3''V_1''} \xrightarrow{u_8} \overline{V_3'V_4'} $\\
$ \overline{V_5'V_3'} \xrightarrow{u_4} \overline{V_1'V_3'} \xrightarrow{u_8^{-1}} \overline{V_4''V_3''} \xrightarrow{u_5^{-1}} \overline{V_5'V_3'} $\\
$ \overline{V_1V_1'} \xrightarrow{u_4^{-1}} \overline{V_5V_5'} \xrightarrow{u_1} \overline{V_1'''V_1''} \xrightarrow{u_3^{-1}} \overline{V_5'''V_5''} \xrightarrow{u_1} \overline{V_1V_1'} $\\
$ \overline{V_5'V_6'} \xrightarrow{u_1} \overline{V_1''V_2''} \xrightarrow{u_8} \overline{V_4'V_2'} \xrightarrow{u_2} \overline{V_4''V_2''} \xrightarrow{u_8} \overline{V_1'V_2'} \xrightarrow{u_1^{-1}} \cdots$ \\
$ \cdots \overline{V_5''V_6''} \xrightarrow{u_5^{-1}} \overline{V_4'V_6'} \xrightarrow{u_2} \overline{V_4''V_6''} \xrightarrow{u_5^{-1}} \overline{V_5'V_6'} $\\
$ \overline{V_1V_2} \xrightarrow{u_1^{-1}} \overline{V_5'''V_6'''} \xrightarrow{u_7^{-1}} \overline{V_4V_6} \xrightarrow{u_2^{-1}} \overline{V_4'''V_6'''} \xrightarrow{u_7^{-1}} \cdots $\\
$ \overline{V_5V_6} \xrightarrow{u_1} \cdots\overline{V_1'''V_2'''} \xrightarrow{u_6} \overline{V_4V_2} \xrightarrow{u_2^{-1}} \overline{V_4'''V_2'''} \xrightarrow{u_6} \overline{V_1V_2} $ \\
\hline
\end{tabular}
\end{center}
\end{table}

\begin{table*}[btp] 
\caption{Action of the generators on the vertices of $\mathcal{P}$.}
\label{action2}
\begin{center}
\begin{tabular}{|l|c|}
\hline
$u_1(S(u_1^{-1}))=S(u_1)$ & $u_1(V_5)=V_1''',\;u_1(V_5')=V_1'',\;u_1(V_5'')=V_1',\;u_1(V_5''')=V_1$,\\
& $u_1(V_6)=V_2''',\; u_1(V_6')=V_2'',\; u_1(V_6'')=V_2',\; u_1(V_6''')=V_1$\\
$u_2(S(u_2^{-1}))=S(u_2)$ & $u_2(V_6')=V_2''),\; u_2(V_4')=V_4''),\; u_2(V_2')=V_6''$,\\
& $u_2(V_6''')=V_2,\; u_2(V_4''')=V_4,\;u_2(V_2''')=V_6$ \\
$u_3(S(u_3^{-1}))=S(u_3)$ & $u_3(V_3'')=V_3'',\;u_3(V_3''')=V_3''',\; u_3(V_5'')=V_1'',\; u_3(V_5''')=V_1'''$\\
$u_4(S(u_4^{-1}))=S(u_4)$ & $u_4(V_3)=V_3,\;u_4(V_3')=V_3',\;u_4(V_5')=V_1',\;u_4(V_5)=V_1$\\
$u_5(S(u_5^{-1}))=S(u_5)$ & $u_5(V_3')=V_3'',\;u_5(V_6')=V_6'',\;u_5(V_5')=V_4'',\;u_5(V_4')=V_5''$\\
$u_6(S(u_6^{-1}))=S(u_6)$ & $u_6(V_3''')=V_3,\; u_6(V_4''')=V_1,\; u_6(V_1''')=V_4,\; u_6(V_2''')=V_2$\\
$u_7(S(u_7^{-1}))=S(u_7)$ & $u_7(V_3)=V_3''',\; u_7(V_5)=V_4''',\;u_7(V_6)=V_6''',\;u_7(V_4)=V_5'''$\\
$u_8(S(u_8^{-1}))=S(u_8)$ & $u_8(V_4'')=V_1',\; u_8(V_2'')=V_2',\;u_8(V_3'')=V_3',\;u_8(V_1'')=V_4''$ \\
\hline
\end{tabular}
\end{center}
\end{table*}

A complete set of relations is listed in Table \ref{relations}. Except for the first four, the products correspond to the identity in $\PSL_2(\C)$ (thus, a trivial rotation). By computing the eigenvalues of $u_3,u_4, u_2u_1$ and $u_2u_1^{-1}$, we find that they are indeed conjugated to rotations of an angle $\frac{2\pi}{3}$ around the axis $\{x=0,y=0\}$.\footnote{ $g \in \SL_2(\C)$ is called \emph{elliptic}, and is a rotation around a fixed geodesic, if and only if $\tr(g) \in \R$ and $\abs{\tr(g)}<2$, see \cite{EGM}, Prop. 1.4. If its eigenvalues are $e^{i\beta},e^{-i\beta}$, then the angle of rotation is $2\beta$.} 
\medskip \par
We have thus shown that $\mathcal{P}$ is a Dirichlet polyhedron for some subgroup $\Gamma$ of $\mo^1$. But if $\Gamma$ were a proper subgroup, the volume of $\mathcal{P}$ would be a multiple of the volume of the fundamental polyhedron for $\mo^1$ that we computed in (\ref{volume}), that is it should be at least $2 \cdot 4.88514\cdots =9.77029\cdots$.\\
So the last step of the proof that $\mathcal{P}$ is a fundamental polyhedron for $\mo^1$ is the following:
\begin{lem} \label{volume_lemma}
$\Vol(\mathcal{P})<9.77029\cdots$.
\end{lem}
The proof of this fact is rather tedious and is reported in the Appendix.

\begin{rem}
From the coordinates of the vertices of $\mathcal{P}$, one finds that the radius of the smallest hyperbolic sphere containing $\mathcal{P}$ is
$$R_{\max}=\arccosh(2.2360\cdots)=1.4436\cdots,$$
while the minimum of the distances between $J$ and the vertices of $\mathcal{P}$ is
$$R_{\min}=\arccosh(1.9069\cdots)=1.2614\cdots$$
\end{rem}

\section{Conclusions}
In this paper we have introduced a right preprocessing method for the decoding of space-time block codes based on quaternion algebras, which allows to improve the performance of suboptimal decoders and reduces the complexity of ML decoders. \\
The new method exploits the algebraic structure of the code, by approximating the channel matrix with a unit in the maximal order of the quaternion algebra. Our simulations show that algebraic reduction and LLL reduction have similar performance. However in the case of slow fading, unlike LLL reduction, algebraic reduction requires only a slight adjustment of the previous approximation at each time block, without needing to perform a full reduction.\\
In future work we will deal with the generalization of algebraic reduction to higher-dimensional space time codes based on cyclic division algebras.

\appendix

\subsection{Proof of Lemma \ref{volume_lemma}} \label{volume_proof}
In order to prove that the volume of $\mathcal{P}$ is smaller than the required constant, we can compute the volume of the hyperbolic polyhedron $\mathcal{Q}$ enclosed by $S(u_1)$, the plane $\{r=-\frac{\bt}{2}\}$, and the spheres $S(u_2)$,$S(u_2^{-1})$,$S(u_1^{-1})$,$S(u_4)$,$S(u_3)$,$S(u_4^{-1})$,$S(u_3^{-1})$. Clearly $\mathcal{Q} \supset \mathcal{P}$.
Recalling the definition of the hyperbolic volume in (\ref{dv}), the volume of the spherical sector $\mathcal{T}$ enclosed by $S(u_1)$ and $\left\{r=-\frac{\bt}{2}\right\}$ is 
\begin{equation*}
\int_{-\frac{\bt}{2}}^{\theta}\frac{\pi(\theta^2-r^2)}{r^3}dr=\pi\left(-\frac{1}{2}-\ln(\theta)+2\theta^4+\ln\left(-\frac{\bt}{2}\right)\right)=36.2937\cdots
\end{equation*}

To this volume we must subtract the volume of the intersection of $\mathcal{T}$ with the chosen spheres. \\
From the expression for the area of the intersection of two circles of radii $R_1$ and $R_2$ whose centers have distance $d$ \cite{We}
{\allowdisplaybreaks
\begin{multline*}
A(R_1,R_2,d)=R_1^2\arccos\left(\frac{d^2+R_1^2-R_2^2}{2dR_1}\right)+R_2^2\arccos\left(\frac{d^2+R_2^2-R_1^2}{2dR_2}\right)+ \\
+\frac{1}{2}\sqrt{(-d+R_1-R_2)}\sqrt{(d+R_1-R_2)(d-R_1+R_2)(d+R_1+R_2)},
\end{multline*}
}%

we obtain the area of the horizontal sections of $\mathcal{T} \cap S(u_2)$. Since $R_1=\sqrt{\theta^2-\bar{r}^2}$, $R_2=\sqrt{1-\bar{r}^2}$ are the radii of $S(u_1) \cap \{r=\bar{r}\}$, $S(u_2) \cap \{r=\bar{r}\}$ respectively, and the distance between the centers is $d=\sqrt{2}$, we find 
\begin{multline*}
A(R_1,R_2,d)=\pi(1-\bar{r}^2)+(\bar{r}^2-1)\arccos\left(\frac{\sqrt{2}}{4}\frac{\theta-2}{\sqrt{1-\bar{r}^2}}\right)+\\
+(\theta^2-\bar{r}^2)\arccos\left(\frac{\sqrt{2}}{4}\frac{1+\theta^2}{\sqrt{\theta^2-\bar{r}^2}}\right)-\frac{1}{2}\sqrt{-1+6\theta^2-\theta^4-8v^2},
\end{multline*}
which is defined for $\bar{r} \leq \frac{\sqrt{9+3\sqrt{5}}}{4}$. In conclusion,
\begin{equation*}
\Vol(\mathcal{T} \cap S(u_2))=\Vol(\mathcal{T} \cap S(u_2^{-1}))=\int_{-\frac{\bt}{2}}^{\frac{\sqrt{9+3\sqrt{5}}}{4}}\frac{A(R_1,R_2,d)}{r^3}dr=5.96793\cdots
\end{equation*}

Proceeding in the same way, one can compute
\begin{align*}
&\Vol(\mathcal{T}\cap S(u_4))=\Vol(\mathcal{T} \cap S(u_3))=5.34536\cdots,\\
&\Vol\left((\mathcal{T} \cap S(u_1^{-1}))\setminus (S(u_1^{-1}) \cap (S(u_2) \cup S(u_2^{-1}))\right)=2.49982\cdots,\\
&\Vol\left((\mathcal{T} \cap S(u_3^{-1}))\setminus (S(u_3^{-1}) \cap (S(u_3) \cup S(u_1^{-1}))\right)=0.70490\cdots
\end{align*}
Therefore the volume of $\mathcal{Q}$ is less than 
\begin{equation*}
36.29366-2\cdot5.96793-2\cdot5.34536-2.49982-2\cdot0.70490=9.75746<9.77029, 
\end{equation*}
which completes our proof. \qed


\begin{thebibliography}{40}
\itemsep=1mm
\parskip=0mm
\bibitem{BRV} J-C. Belfiore, G. Rekaya, E. Viterbo, \vv{The Golden Code: a $2 \times 2$ full-rate Space-Time Code with non-vanishing determinants}, \emph{IEEE Trans. Inform. Theory}, vol 51 n.4, 2005
\bibitem{CJLdR} C. Corrales, E. Jespers, G. Leal, A. del Rio, \vv{Presentations of the unit group of an order in a non-split quaternion algebra}, \emph{Advances in Mathematics}, 186 n. 2 (2004) 498--524
\bibitem{DEC} M. O. Damen, H. El Gamal, G. Caire, \vv{MMSE-GDFE lattice decoding for solving under-determined linear systems with integer unknowns}, Proceedings of ISIT 2004, Chicago, USA 
\bibitem{Ed} A. Edelman, \vv{Eigenvalues and condition numbers of random matrices}, Ph.D. Thesis, MIT 1989
\bibitem{EGM} J. Elstrodt, F. Grunewald, J. Mennicke, \vv{Groups acting on Hyperbolic Space}, Springer Monographs in Mathematics, 1998
\bibitem{EP} D. Epstein, C. Petronio, \vv{An exposition of Poincaré's Polyhedron Theorem}, \emph{L'Enseignement Ma\-thé\-ma\-tique}, 40 (1994), 113--170
\bibitem{Go} N. R. Goodman, \vv{The distribution of the determinant of a complex Wishart distributed matrix}, \emph{Ann. Math. Statist.}, 34, 178-180
\bibitem{HLRV} C. Hollanti, J. Lahtonen, K. Ranto, R. Vehkalahti, \emph{On the densest MIMO lattices from cyclic division algebras}, submitted to \emph{IEEE Trans. Inform. Theory}
\bibitem{Kl} E. Kleinert, \vv{Units of classical orders: a survey}, \emph{L'Enseignement Mathématique}, 40 (1994), 205--248
\bibitem{La} S. Lang, \vv{Algebraic number theory}, Springer-Verlag 2000
\bibitem{LRBV} L. Luzzi, G. Rekaya Ben-Othman, J.-C. Belfiore, E. Viterbo, \emph{Golden Space-Time Block} \emph{Coded Modulation}, submitted to \emph{IEEE Trans. Inform. Theory}
\bibitem{MEDC} A. D. Murugan, H. El Gamal, M. O. Damen, G. Caire, \vv{A unified framework for tree search decoding: rediscovering the sequential decoder}, \emph{IEEE Trans. Inform. Theory}, vol 52 n. 3, 2006
\bibitem{MR} C. Maclachlan, A. W. Reid, \vv{The arithmetic of hyperbolic 3-manifolds}, Graduate texts in Mathematics, Springer, 2003
\bibitem{ORBV} F. Oggier, G. Rekaya, J.-C. Belfiore, E. Viterbo, \vv{Perfect Space-Time Block Codes}, \emph{IEEE Trans. Inform. Theory}, vol. 52 n.9, 2006 
\bibitem{Pr} J. Proakis, \vv{Digital communications}, McGraw-Hill 2001
\bibitem{RBV} G. Rekaya, J-C. Belfiore, E. Viterbo, \vv{A very efficient lattice reduction tool on fast fading channels}, Proceedings of ISITA 2004, Parma, Italy
\bibitem{SRS} B. A. Sethuraman, B. Sundar Rajan, V. Shashidar, \vv{Full-diversity, high-rate space-time block codes from division algebras}, \emph{IEEE Trans. Inform. Theory}, vol 49 n. 10, 2003, pp. 2596--2616
\bibitem{Sw} R. G. Swan, \vv{Generators and relations for certain special linear groups}, \emph{Adv. Math.} 6, 1--77
\bibitem{TMK} M. Taherzadeh, A. Mobasher, A. K. Khandani, \vv{LLL reduction achieves the receive diversity in MIMO decoding}, \emph{IEEE Trans. Inform. Theory}, vol 53 n. 12, 2007, pp 4801--4805
\bibitem{Vig} M-F. Vignéras, \vv{Arithmétique des Algèbres de Quaternions}, Lecture Notes in Mathematics, Springer Verlag 1980
\bibitem{We} Weisstein, Eric W. \vv{Circle-Circle Intersection}, from \emph{MathWorld-}\emph{A Wolfram Web Resource.} \texttt{http:}\texttt{//mathworld.wolfram.com/}\texttt{Circle-}\-\texttt{CircleIntersection.html}
\end{thebibliography}
\end{document}